\documentclass{aa}  

\usepackage{graphicx}	
\usepackage{amsmath}	
\usepackage{amsmath, amsfonts, amssymb, graphics,graphicx, wrapfig, nicefrac}
\usepackage{multirow}
\usepackage{multicol}
\usepackage{graphicx}
\usepackage{rotating}
\usepackage{mathrsfs}
\usepackage{float}
\graphicspath{{./}{figures/}}
\usepackage{txfonts}
\usepackage[colorlinks=true,linkcolor=blue, citecolor=blue,urlcolor=magenta]{hyperref}
\begin{document}

   \title{The evolution of CH in {\it Planck} Galactic Cold Clumps}


   \author{Gan Luo
          \inst{1}
          \and
          Arshia M. Jacob\inst{2,3}
          \and
          Marco Padovani\inst{4}
          \and
          Daniele Galli\inst{4}
          \and
          Ana L\'opez-Sepulcre\inst{5,1}
          \and
          Ningyu Tang\inst{6}
          \and
          Di Li\inst{7,8}
          \and
          Jing Zhou\inst{9,10}
          \and
          Pei Zuo\inst{8}
          }

   \institute{Institut de Radioastronomie Millimetrique, 300 rue de la Piscine, 38400, Saint-Martin d’Hères, France\\
            \email{luo@iram.fr}
        \and
            I.Physikalisches Institut, Universitätzu Köln, Zülpicher Str.77, 50937 Köln, Germany
        \and
            Max-Planck-Institut für Radioastronomie, Auf dem Hügel 69, 53121 Bonn, Germany
        \and
            INAF-Osservatorio Astrofisico di Arcetri, Largo E. Fermi 5, 50125 Firenze, Italy
        \and
            Univ. Grenoble Alpes, CNRS, IPAG, 38000 Grenoble, France
        \and
            Department of Physics, Anhui Normal University, Wuhu, Anhui 241002, China
        \and
             Department of Astronomy, Tsinghua University, Beijing 100084, China
        \and
             CAS Key Laboratory of FAST, National Astronomical Observatories, Chinese Academy of Sciences, Beijing 100101, China
        \and
            Space Engineering University, Beijing 101416, China
        \and
            School of Astronomy and Space Science, Nanjing University, Nanjing 210093, China
             }

   \date{Received xx; accepted xx}

\abstract{Methylidyne (CH) has long been considered a reliable tracer of molecular gas in the low-to-intermediate extinction range. Although extended CH 3.3 GHz emission is commonly observed in diffuse and translucent clouds, observations in cold, dense clumps are rare. In this work, we conducted high-sensitivity CH observations toward 27 Planck Galactic Cold Clumps (PGCCs) with the Arecibo 305m telescope. Toward each source, the CH data were analyzed in conjunction with $^{13}$CO (1--0) emission, H\,{\sc i} narrow self-absorption (HINSA), and H$_2$ column densities inferred from thermal dust emission. Our results revealed ubiquitous subsonic velocity dispersions of CH, in contrast to $^{13}$CO, which is predominantly supersonic. The findings suggest that subsonic CH emissions may trace dense, low-turbulent gas structures in PGCCs. To investigate how environmental parameters, particularly the cosmic-ray ionization rate (CRIR), affect the evolution of CH in PGCCs, we estimated upper limits for the CRIR using HINSA. The derived values span $(8.1\pm4.7)\times10^{-18}$ to $(2.0\pm0.8)\times10^{-16}$ s$^{-1}$ over an H$_2$ column density range of $(1.7\pm0.2)\times10^{21}$ to $(3.6\pm0.4)\times10^{22}$~cm$^{-2}$. This result favors theoretical predictions of a cosmic-ray attenuation model, in which the interstellar spectra of low-energy CR protons and electrons match {\it Voyager} measurements, although alternative models cannot yet be ruled out. The abundance of CH decreases with increasing column density, while showing a positive dependence on the CRIR, which requires atomic oxygen not heavily depleted to dominate CH destruction in PGCCs. By fitting the abundance of CH with an analytic formula, we place constraints on atomic O abundance ($2.4\pm0.4\times10^{-4}$ with respect to H column density) and C$^+$ abundance ($7.4\pm0.7\times10^{13}\zeta_2/n_{\rm H_2}$). These findings indicate that CH formation is closely linked to the C$^+$ abundance, regulated by cosmic-ray ionization, while other processes, such as turbulent diffusive transport, might also contribute a non-negligible effect to CH formation.}
 


   \maketitle
%

\section{Introduction}

Molecular gas (predominantly H$_2$) is the raw material for forming stars. However, cold molecular clouds ($T \lesssim 15$~K) cannot be directly probed through observations of H$_2$ due to the large energy level spacing \citep[the upper-level energies of the two lowest rotational transitions are $E/k\approx 510$~K and 1015~K above the ground level,][]{Shull1982}. Moreover, H$_2$ does not possess a permanent dipole moment and only emits weakly via its quadrupole transitions, which are presently accessible with the capability of the {\it James Webb} Space Telescope \citep{Bialy2022,Padovani2022}. Thus, it is necessary to study the physical and chemical evolution of molecular clouds through molecular line transitions of other interstellar species. 
As one of the first molecules detected in the interstellar medium (ISM) through optical absorption lines \citep{Dunham1937,Swings1937}, methylidyne (CH) has been observed in various environments, from diffuse molecular clouds to massive star-forming regions \citep{Rydbeck1976,Genzel1979,Magnani1989,Sheffer2008,Jacob2021}. It is of particular interest because of its fairly constant abundance with respect to H$_2$ ($\sim4\times10^{-8}$) in diffuse and translucent clouds \citep[$A_{\rm V} \leq$ 2 mag,][]{Snow2006} owing to which it has been established as an important tracer for molecular gas in these environments \citep{Federman1982,Liszt2002,Sheffer2008}. In addition, CH has been ubiquitously observed through weak maser emission arising from hyperfine structure transitions between its ground state $\Lambda$-doublet at 3.3 GHz \citep[e.g., quasars, H\,{\sc ii} regions,][]{Rydbeck1976,Genzel1979,Jacob2021,Tang2021}. 

CH is one of the most important precursors for complex molecules \citep{Van1988,Bialy2015}. In diffuse clouds, CH is primarily formed through the dissociative recombination of CH$_2^+$ and CH$_3^+$ with electrons, where the latter is formed via the reaction of H$_2$ and CH$^+$ \citep{Black1973,Bialy2015,Bisbas2019,Jacob2023}. However, due to the highly endothermic reaction that forms CH$^+$ (requiring energies of 4640~K), the high CH$^+$ abundances observed in diffuse molecular clouds have been a longstanding puzzle \citep[see e.g.,][]{Federman1996,Godard2009,Godard2014,Godard2023}. 
The spatial distribution of CH extends into the outskirts of molecular clouds, where CO emission is faint or absent, and the linewidth of CH is observed to be broader than that of CO \citep{Magnani1989,Magnani1993,Sheffer2008,Xu2016}. Thus, CH has long been proposed to be a reliable proxy for H$_2$ in low-to-intermediate density clouds, particularly in regions where the H\,{\sc i}-to-H$_2$ phase transition is expected to occur \citep{Sheffer2008,Sakai2012,Xu2016,Gerin2016,Jacob2019}. 

In dense environments, CH 3.3 GHz transitions have predominantly been observed toward massive star-forming regions (e.g., Sgr B2), which are strong toward bright continuum background sources with strong far-infrared (FIR) radiation \citep{Genzel1979,Jacob2021}. In such environments, the emission in the lower satellite line (3264 MHz) is usually enhanced due to pumping effects from the radiative trapping caused by the overlap of CH's rotational transitions at sub-millimeter and FIR wavelengths \citep{Jacob2024}. 
Measurements of CH in cold, dense clumps remain limited. In contrast to the ubiquitous broad linewidths in diffuse molecular clouds and massive star-forming regions, CH emission toward the TMC-1 region exhibits a combination of broad and narrow velocity components \citep{Sakai2012}. While the extended broad component over the whole region mainly traces the envelope of the cloud, the narrow components coincide with C$^{18}$O peaks, suggesting that they may trace gravitationally bound dense cores rather than transient coherent structures \citep{Sakai2012}. However, due to limitations in spatial resolution and the small sample size of existing observations, the physical origin of these narrow linewidth components -- specifically, whether they trace dense cores or low-turbulent coherent structures in cold clumps -- remains uncertain.

In this work, we conduct CH 3.3 GHz observations toward 27 Planck Galactic Cold Clumps (PGCCs) with the Arecibo 305m telescope. These sources are selected from nearby high-latitude low-mass star-forming clouds (e.g., Taurus, Perseus). The dust temperatures ($T_{\rm d}$) of the selected sources range from 10 to 13~K and the total H volume densities ($n_{\rm H}$\footnote{For clarification, we use $n_{\rm H}$, $n_{\rm H_2}$, and $n_{\rm HI}$ to represent the total H, H$_2$, and H\,{\sc i} volume densities throughout the text. In PGCCs, $n_{\rm H} = n_{\rm HI} + 2n_{\rm H_2} \approx 2n_{\rm H_2}$.}) range from $(1.2\pm0.5)\times10^3$ to $(2.8\pm2.5)\times10^4$~cm$^{-3}$ \citep{Planck2016}, representing cold and translucent-dense environments. 

This paper is organized as follows: the observations and archival data used in this work are presented in Section \ref{sec:obs}. In Section \ref{sec:results}, we present the methods used to fit the spectra and derive the column densities. In Section \ref{sec:discussion}, we discuss the uncertainties in deriving the molecular column densities, the molecular abundances, the estimation of cosmic-ray ionization rates, and the evolution of CH in PGCCs. The main results and conclusions are summarized in Section \ref{sec:conclusion}.

\section{Observations}\label{sec:obs}

\subsection{Arecibo CH observations}\label{sec:obs_ch}

We have observed the three hyperfine structure lines of CH within the ground state $\Lambda$-doublet $^2\Pi_{1/2}$, $J$ = 1/2 energy level \citep[3263.794, 3335.479, and 3349.193 MHz,][]{Truppe2013} toward 27 PGCCs, using the S-band high receiver with the total power ON mode during September 1-11, 2018 (Project ID: A3224). The three CH transitions were placed into three sub-bands, each with a bandwidth of 3.125 MHz and a spectral resolution of 381 Hz (corresponding to $\sim$0.034 km\,s$^{-1}$ at 3335 MHz). The final spectra are smoothed to a velocity resolution of $\sim$0.14 km\,s$^{-1}$ to obtain a better S/N. The integration time was 20 mins for each source, resulting in an rms noise level of 0.012$\sim$0.020~K per 0.14 km\,s$^{-1}$. 
The beam size of CH at 3335 MHz is $\sim$1.7$'$ and the beam efficiency is 40$\%$. A complete list of the observed sources is shown in Table \ref{table:tab1}.

\subsection{$^{13}$CO (1--0) observations}\label{sec:obs_13co}

The majority of the $^{13}$CO (1--0) data (hereafter, $^{13}$CO) were taken from the 100 deg$^2$ survey of the Taurus molecular cloud \citep{Goldsmith2008} and the COMPLETE survey of the Perseus molecular cloud \citep{Ridge2006}, both observed with the FCRAO 13.7-m telescope using the on-the-fly (OTF) mapping mode and have a beam size of $\sim$46$''$ at 110 GHz. The velocity resolutions for the data in Taurus and Perseus are 0.27 km\,s$^{-1}$ and 0.066 km\,s$^{-1}$, and the mean rms in antenna temperature ($T_{\rm A}^*$) per velocity channel are 0.125 and 0.17~K, respectively. The main beam efficiency is 0.49 at 110\,GHz. 

The above surveys did not cover seven sources in our sample, for which we used the Delingha 13.7-m telescope with OTF mode during October 11-17, 2024 (Project ID: 24B003) to observe $^{13}$CO (1--0). Each source covers a 5$'$ $\times$ 5$'$ region. The spectral resolution is 0.083 km\,s$^{-1}$ and resultant rms on main-beam temperature scales ($T_{\rm mb}$) per velocity channel is 0.11~K. The final map of $^{13}$CO was smoothed to the same angular resolution as that of CH.

\subsection{Archival H\,{\sc i} data}\label{sec:obs_hi}

We used H\,{\sc i} data from the Galactic Arecibo L-Band Feed Array H\,{\sc i} (GALFA-H\,{\sc i}) survey \citep{Peek2011,Peek2018}. The survey covers a large area (13000 $\mathrm{deg^2}$) with high spatial resolution ($\sim$4$'$), high spectral resolution (0.18\,km\,s$^{-1}$), and large bandwidth ($-$700\,km\,s$^{-1}$\ $<$ $\mathrm{V_{LSR}}$ $<$ +700\,km\,s$^{-1}$). The datacube was regridded to a spatial resolution of 1$'$), and the typical noise level is 80\,mK per 1\,km\,s$^{-1}$ channel.

\subsection{Archival continuum data}\label{sec:obs_continuum}

We used dust thermal emission to infer the dust temperature ($T_{\rm d}$) and column density of H$_2$ ($N_{\rm H_2}$) toward each source. The available maps of $T_{\rm d}$ and $N_{\rm H_2}$ toward Taurus and Perseus were taken from the {\it Herschel} Gould Belt Survey  \citep[HGBS;][]{Andre2010,Palmeirim2013,Marsh2016,Pezzuto2021,Kirk2024}. The final products have an angular resolution of 36.3$''$, but they are smoothed to the same angular resolution as CH. We consider the $N_{\rm H_2}$ values to have an uncertainty of 10\% \citep{Roy2014}. There are 11 sources without available {\it Herschel} observations, for which we used the $T_{\rm d}$ and $N_{\rm H_2}$ values obtained from {\it Planck} \citep{Planck2016}. For each source, we obtained $n_{\rm H}$ from {\it Planck}, in which the value is derived by assuming the pathlength to be the same as the diameter of the clump. The values of $n_{\rm H}$, $T_{\rm d}$, and $N_{\rm H_2}$ toward all sources are listed in Table \ref{table:tab1}. 

\section{Results}\label{sec:results}
\subsection{Spectra line fitting}\label{sec:fit}

\begin{figure*}
\centering
\includegraphics[width=0.95\linewidth]{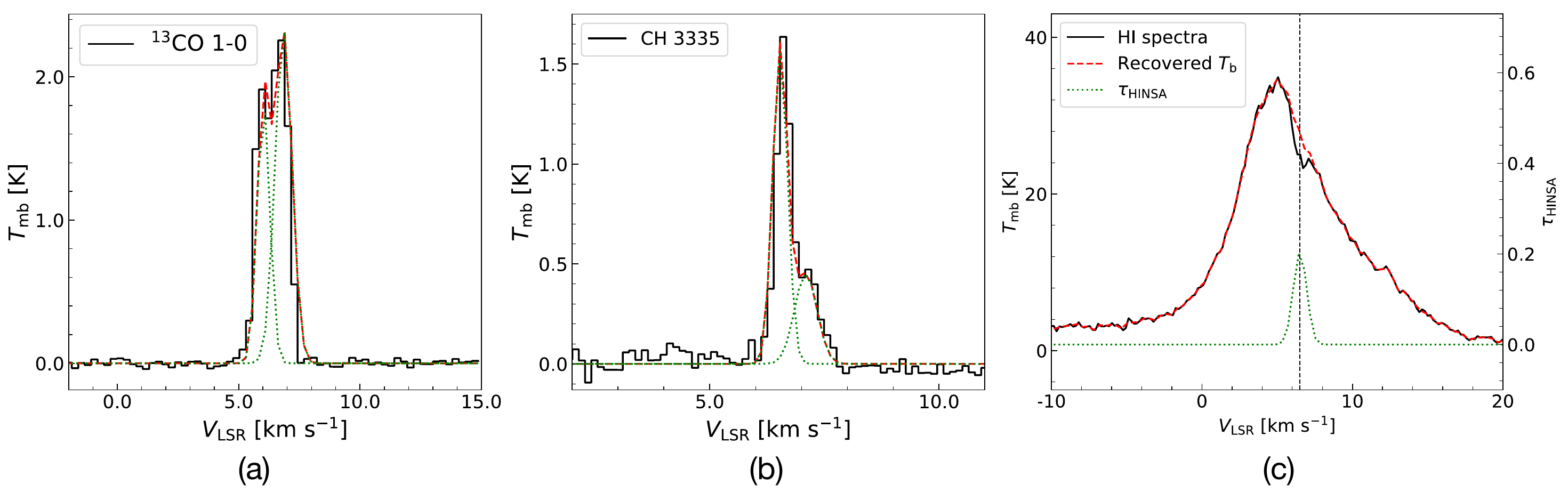}
\caption{Example presenting the (a): $^{13}$CO, (b) CH 3335~MHz, and (c) H\,{\sc i} spectra toward G168.13-16.39. In subplots (a) and (b), the red dashed curve represents the Gaussian fitting results, while the green dotted curves represent each component. In subplot (c), the red dashed curve denotes the recovered background H\,{\sc i} emission without absorption ($T_{\rm b}$). The green dotted and black vertical lines denote the decomposed $\tau$ and $V_{\rm lsr}$ of HINSA, respectively. \label{fig:line_fit}}
\end{figure*}

The spectra of both $^{13}$CO and CH ground state hyperfine structure lines are fitted with Gaussian profiles. For sources with multiple peaks in the line profile, we used multiple Gaussian components. We performed Gaussian decomposition using the {\it curve\_fit} package in {\it scipy} with the Levenberg-Marquardt algorithm. The resultant local standard of rest velocity ($V_{\rm lsr}$), linewidth ($\Delta V$), and $T_{\rm mb}$ for each component are listed in Tables \ref{table:tab2} and \ref{table:tab3}. The statistic from the Gaussian fitting shows that the hyperfine structure line ratio of CH 3335 and CH 3264 is 1.9$\pm$0.4, and the ratio of CH 3349 and CH 3264 is 1.0$\pm$0.2, consistent with the intrinsic line ratios under conditions of local thermodynamic equilibrium. An example of the fitting profile toward G168.13-16.39 is shown in Fig.~\ref{fig:line_fit}(a) and (b). 

\begin{figure*}
\centering
\includegraphics[width=0.95\linewidth]{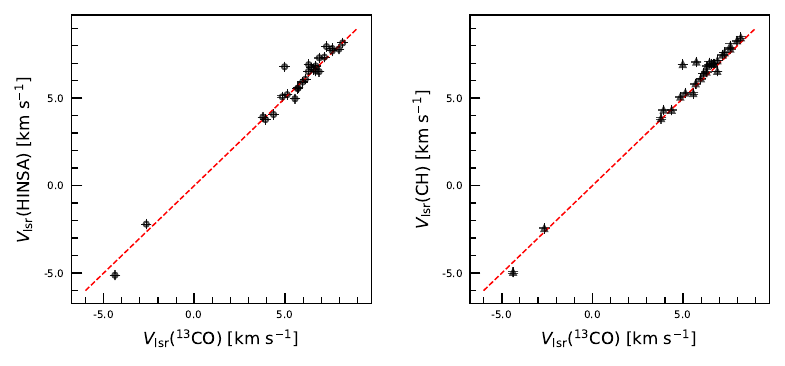}
\caption{Comparison between the $V_{\rm lsr}$'s of $^{13}$CO emission with that of HINSA (left) and CH (right). The red dashed lines mark a slope of unity. \label{fig:Vlsr}}
\end{figure*}

For the H\,{\sc i} spectra, we only focus on the H\,{\sc i} narrow self-absorption (HINSA) profile, which defines the cold H\,{\sc i} that is cooled down by collisions with cold H$_2$ inside molecular clouds \citep{Li2003,Goldsmith2005}. We used the second-order derivative method \citep[described in][]{Krco2008} to fit the line profile and obtain the optical depth of the HINSA features. The assumptions and modeling of the recovered background H\,{\sc i} emission follow the same steps as in \citet{Luo2024a}. We used the Markov Chain Monte Carlo (MCMC) method within the $emcee$ code \citep{Foreman-Mackey2013} to sample the free parameters (optical depth $\tau$ of HINSA, $V_{\rm lsr}$, and $\Delta V$) and the posterior probability distribution. The prior function is set to reject sampling outside the defined parameter space: 1) $V_{\rm lsr}$ within $\pm$0.5 km\,s$^{-1}$ offset with respect to the $V_{\rm lsr}$ of $^{13}$CO and CH, and 2) $\Delta V$ between 0.35 to 3 km\,s$^{-1}$\footnote{The pure thermal linewidth ($\Delta V_{\rm therm}$) for H\,{\sc i} at 2.73~K is 0.35 km\,s$^{-1}$, we would not expect the linewidth below this value. The largest linewidth of $^{13}$CO in our sample is $2.33\pm0.08$ km\,s$^{-1}$. The HINSA signature, by its nature, would not exhibit significantly larger linewidth than molecular lines.}. The MCMC fitting shows good convergence in the sampled parameter space for most sources (see Fig.~\ref{fig:hinsa_fit} for an example). And the fitting values of $V_{\rm lsr}$ between HINSA and $^{13}$CO or CH do not show significant differences (Fig.~\ref{fig:Vlsr}). An example of the HINSA fitting toward G168.13-16.39 is shown in Fig.~\ref{fig:line_fit}(c). The fitting results of HINSA are listed in Table \ref{table:tab4}. Spectra toward all sources can be found in Appendix \ref{sec:lines}. 

\subsection{Non-thermal velocity dispersion}\label{sec:linewidth}

\begin{figure*}
\centering
\includegraphics[width=0.95\linewidth]{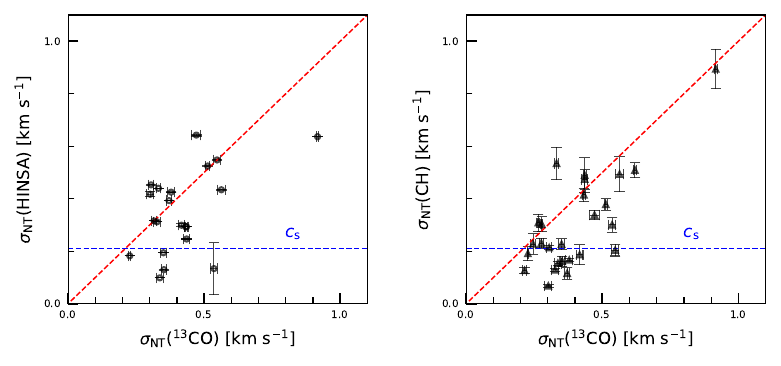}
\caption{Comparison between the non-thermal velocity dispersion ($\sigma_{\rm NT}$) of $^{13}$CO with that of HINSA (left) and CH (right). The red dashed lines mark a slope of unity, blue dashed lines mark the sound speed ($c_{\rm s}$ = 0.21 km\,s$^{-1}$) at $T_{\rm k}$ = 12~K. \label{fig:sigma_NT}}
\end{figure*}

Non-thermal velocity dispersion ($\sigma_{\rm NT}$) describes the random motion of the gas that is not due to thermal motion of particles (e.g., turbulence, bulk motion). We calculate the non-thermal velocity dispersion of each tracer with the following equation:
\begin{equation}
    \sigma_{\rm NT} = \sqrt{\left(\frac{\Delta V}{2\sqrt{\ln 2}}\right)^2-\left(\frac{kT}{m_i}\right)^2},
\end{equation}
where $k$ is the Boltzmann constant, $m_i$ is the mass of a given species, and $T$ is the temperature of the gas, which we assume to be equal to $T_{\rm d}$. Figure \ref{fig:sigma_NT} shows the comparison between the non-thermal velocity dispersion of $^{13}$CO, CH, and HINSA\footnote{When there is more than one velocity component, we compare the strongest components.}. 

For HINSA calculation, there are 5 sources with velocity dispersion lower than the thermal velocity dispersion of H\,{\sc i} ($\sigma_{\rm therm} \sim 0.31$ km\,s$^{-1}$ at 12~K), meaning that the kinetic temperature of HINSA must be lower than $T_{\rm d}$. These sources presumably have subsonic non-thermal velocity dispersion. However, without additional information on the actual kinetic temperature, we cannot obtain accurate non-thermal velocity dispersion. Two sources that did not show clear HINSA signatures were excluded from the related analysis. The values of HINSA $\sigma_{\rm NT}$ are in the range of 0.10 to 0.64~km\,s$^{-1}$, in which 5 sources have subsonic velocity dispersion ($\sigma_{\rm NT} < c_{\rm s}$, where $c_{\rm s} = \sqrt{kT/\mu m_{\rm H}}=0.21$~km\,s$^{-1}$ is the sound speed at $T = 12$~K and $\mu$ = 2.3 is the mean molecular weight.). 

The non-thermal velocity dispersions of $^{13}$CO range from 0.215$\pm$0.007 to 0.917$\pm$0.005~km\,s$^{-1}$. One source shows transonic non-thermal velocity dispersion ($\sigma_{\rm NT} \approx c_{\rm s}$), while all others are supersonic. Most of the $^{13}$CO components have comparable or larger non-thermal velocity dispersion than that of HINSA.

The non-thermal velocity dispersions of CH range from 0.07$\pm$0.04 to $1.23\pm 0.60$~km\,s$^{-1}$, and is systematically smaller than that of $^{13}$CO. In particular, ten sources show subsonic velocity dispersion and four are transonic ($\sigma_{\rm NT} \approx c_{\rm s}$), whereas almost all $^{13}$CO lines are supersonic ($\sigma_{\rm NT} > c_{\rm s}$). 

\subsection{Column densities of $^{13}$CO}\label{sec:col_13co}

To calculate the column density of $^{13}$CO, we assume that the line optical depth is not far greater than unity \citep[namely, adding an optical depth correction factor to the optically thin assumption,][]{Goldsmith1999}. A convenient expression for computing $^{13}$CO column densities is \citep{Mangum2015,Luo2023a}:
\begin{equation}
\begin{matrix}
N_{\rm ^{13}CO} = 6.57 \times 10^{14} \frac{\tau_{13}}{1-e^{-\tau_{13}}} Q_{\rm rot} \left( 1-e^{-\frac{5.288}{T_{\rm ex}}}\right)^{-1} \frac{\int T_{\rm mb} d\upsilon}{J(T_{\rm ex})-0.89} \;{\rm cm^{-2}} \\
\end{matrix}
\end{equation}
where $T_{\rm ex}$ is the excitation temperature, $Q_{\rm tot}$ is the partition function, $T_{\rm mb}$ is the main beam temperature of $^{13}$CO (1-0) emission, and $J(T_{\rm ex})$ = $(h\nu/k)/[\exp(h\nu/kT)-1]$ is the Rayleigh Jeans equivalent temperature. The optical depth of $^{13}$CO ($\tau_{13}$) is:
\begin{equation}
\tau_{13} = -\ln \left[ 1 - \frac{T_{\rm mb}}{J(T_{\rm ex})-0.89} \right].
\label{eq:tau13}
\end{equation}
The partition function $Q_{\rm rot}$ for linear molecules can be approximately expressed as \citep{McDowell1987}:
\begin{equation}
Q_{\rm rot} = \frac{kT}{hB_0} e^{\frac{hB_0}{3kT}} ,
\label{eq:q_rot}
\end{equation}
where $h$ is the Planck constant, and $B_0$ is the rigid rotor rotation constant of $^{13}$CO.

In our calculation, we assume $T_{\rm ex}$ = $T_{\rm d}$ for $^{13}$CO since it can be easily thermalized in the average environment of PGCCs. The derived values of $N_{\rm ^{13}CO}$ and $\tau_{13}$ for $^{13}$CO are listed in Table \ref{table:tab2}. The column densities of $^{13}$CO range from $(1.5\pm 0.9)\times10^{14}$ to $(1.20\pm 0.07)\times10^{16}$~cm$^{-2}$ and the derived $\tau_{13}$ ranges from $0.04 \pm 0.01$ to $2.0\pm 0.2$. Most of the values of $\tau_{13}$ are below unity, indicating that the non-optically thick assumption is reasonable. 

\subsection{{\sc radex} modeling of CH}\label{sec:col_ch}

The values of $T_{\rm ex}$ for CH are usually adopted as $-60$~K or $-15$~K in the literature \citep{Rydbeck1976,Genzel1979}. However, recent observations and models found that $T_{\rm ex}$ could be above $-1$~K \citep{Dailey2020,Jacob2021,Tang2021}. Using the canonical value of $T_{\rm ex}$ to calculate the column density of CH ($N_{\rm CH}$) would lead to an overestimation by up to an order of magnitude. Since there is no constraint for $T_{\rm ex}$ in cold, dense environments, we use the radiative transfer code {\sc radex} \citep{Van2007} to obtain the column densities of CH and evaluate the potential uncertainties. 

\begin{figure*}
\centering
\includegraphics[width=0.95\linewidth]{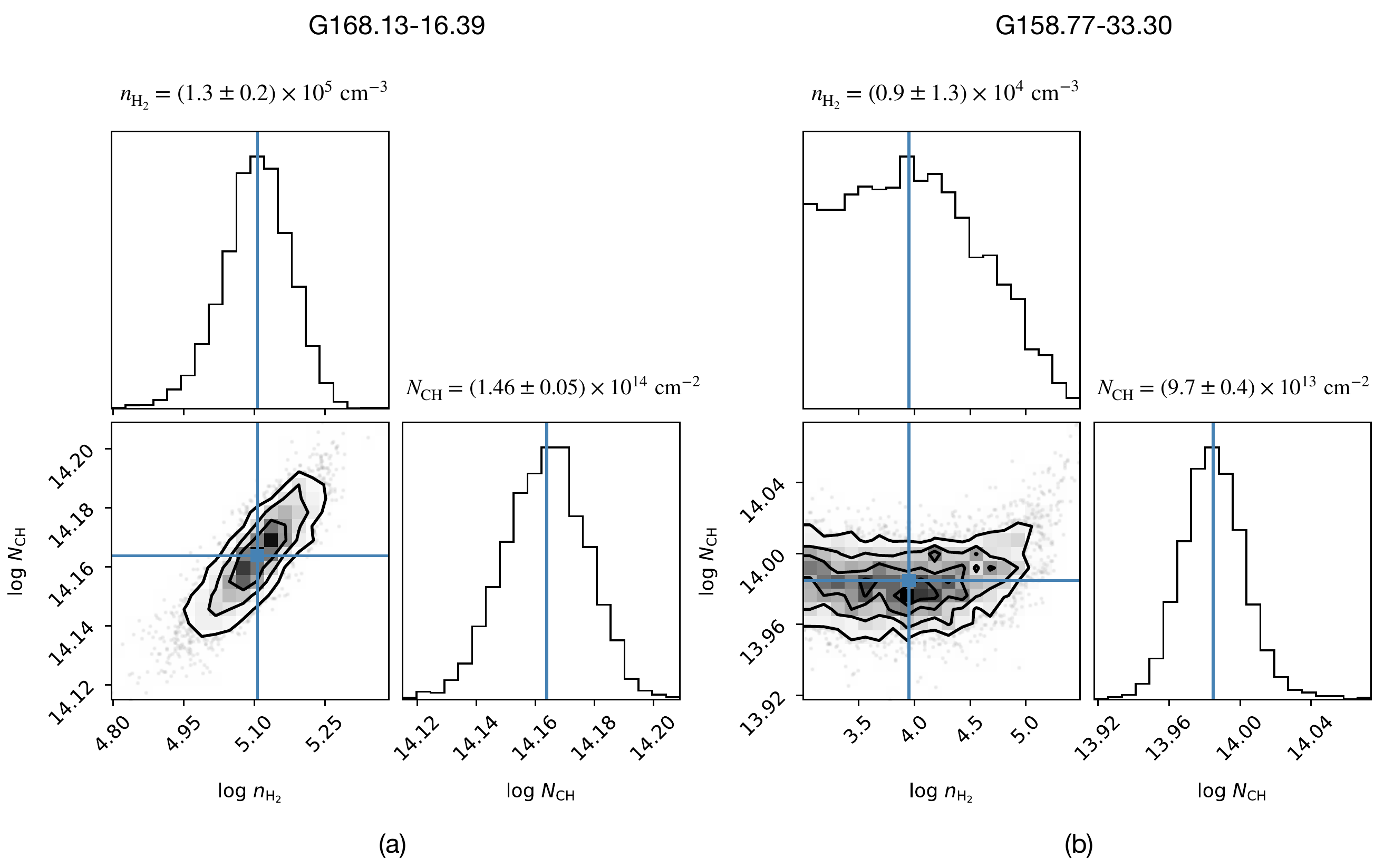}
\caption{The posterior probability distribution of $n_{\rm H_2}$ and the column density of CH ($N_{\rm CH}$) toward (a) G168.13-16.39 at $V_{\rm lsr} = 6.53$ km\,s$^{-1}$ and (b) G158.77-33.30 at $V_{\rm lsr} = -2.44$ km\,s$^{-1}$. The kinetic temperatures used in the models are 11.6~K for G168.13-16.39 and 12.5~K for G158.77-33.30. \label{fig:ch_mcmc}}
\end{figure*}

We use the same treatment for the collision rate coefficients between CH, H$_2$, and H atom as \citet{Jacob2021}. In the fitting procedures, we fix the kinetic temperature ($T_{\rm kin}$) to the same value as $T_{\rm d}$, and set the H$_2$ volume density ($n_{\rm H_2}$) and $N_{\rm CH}$ as free parameters for each velocity component\footnote{The H\,{\sc i} volume density ($n_{\rm HI}$) in the model is scaled according to HINSA abundance (Sect.~\ref{sec:abundances}).}. Figure \ref{fig:ch_mcmc} shows an example of the posterior probability distribution of the free parameters toward G168.13-16.39 and G158.77-33.30. While the modeling toward G168.13-16.39 shows a good convergence, those toward G158.77-33.30 do not well constrain $n_{\rm H_2}$. Nevertheless, $N_{\rm CH}$ does not show large variations even with large uncertainties on $n_{\rm H_2}$\footnote{As long as $T_{\rm ex}$ remains much lower than $-1$~K, the specific choice of $T_{\rm ex}$ does not introduce substantial differences on $N_{\rm CH}$. The median value of $T_{\rm ex}$ in our modeling is $-6.5$~K (Sect.~\ref{sec:excitation}).}. The fitting results for CH toward all sources are listed in Table \ref{table:tab3}. The column densities of CH range from $(4.0\pm3.0)\times10^{12}$ to $(1.8\pm0.1)\times10^{14}$~cm$^{-2}$. 

\subsection{Column densities of HINSA}\label{sec:col_hinsa}

The column density of HINSA ($N_{\rm HINSA}$) can be written as a function of $\tau$ \citep{Li2003}:
\begin{equation}
    N_{\rm HINSA} = 1.95\times10^{18} \tau \ T_{\rm k} \Delta V \; {\rm cm^{-2}},
\end{equation}
where the kinetic temperature of H\,{\sc i}, $T_{\rm k}$ is assumed to be equal to $T_{\rm d}$. The values of $N_{\rm HINSA}$ range from $5.59\times10^{17}$ to $1.61\times10^{19}$~cm$^{-2}$. We adopt an additional uncertainty of 20\% induced by the assumptions of the HINSA fitting (Sect.~\ref{sec:errors}). All the results of HINSA are listed in Table \ref{table:tab4}.

\section{Analysis and discussion}
\label{sec:discussion}

\subsection{Uncertainties}\label{sec:errors}

\begin{figure}
\centering
\includegraphics[width=0.95\linewidth]{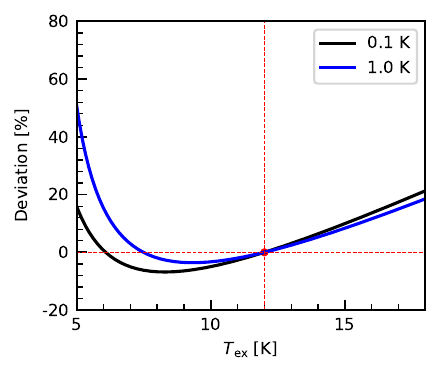}
\caption{Deviations in $N_{\rm ^{13}CO}$ computed assuming $T_{\rm ex} = 12$~K for $T_{\rm mb}$ = 0.1 (in black) and 1~K (in blue). \label{fig:13co_tex}}
\end{figure}

The assumptions we made when calculating the column densities would bring additional uncertainties. When calculating the column density of $^{13}$CO, we assumed $T_{\rm ex} = T_{\rm d}$. However, the value of $T_{\rm ex}$ could be lower than $T_{\rm d}$ when the gas density is low (e.g., $n_{\rm H} \lesssim 10^3$~cm$^{-3}$). Figure \ref{fig:13co_tex} shows the deviation of $^{13}$CO column density with respect to the values at $T_{\rm ex} = 12$~K (median value in our sample) as a function of $T_{\rm ex}$. Even if we consider a variation of $T_{\rm ex}$ from 5 to 20~K, the deviation of $N_{\rm ^{13}CO}$ is less than 50\% for $T_{\rm mb}$ between 0.1 to 1~K\footnote{A source exhibiting strong $^{13}$CO emission is unlikely to exhibit sub-thermal excitation of $^{13}$CO.}. On the other hand, considering $^{12}$CO is always optically thick in these environments, $T_{\rm ex}$ can be roughly estimated through the intensity of $^{12}$CO. The lowest value of the $^{12}$CO peak brightness temperature, $T_{\rm mb}$, in our samples as per archival data from FCRAO is 5.5~K \citep{Goldsmith2008}. Therefore, the excitation temperature should be $T_{\rm ex} \geq 9$~K and the deviation should be within 20\% for all sources. 

The uncertainties in the HINSA column densities are mainly due to the assumed fraction $p$ of H\,{\sc i} in the background and the uncertainty on $T_{\rm k}$ \citep{Li2003}. The former would contribute $\sim 15$\% uncertainty if the majority of H\,{\sc i} is in the background ($p\sim 1$) or $p = 0.83$ \citep[median value in nearby clouds reported by][]{Li2003}. We do not have any independent measurements or inferences of $T_{\rm k}$. In a few cases, $T_{\rm k}$ must be lower than $T_{\rm d}$ since the derived linewidth of HINSA is smaller than the thermal linewidth at $T_{\rm d}$ (see Sect.~\ref{sec:linewidth}). Considering the smallest linewidth of HINSA we obtained ($\Delta V = 0.58$\, km\,s$^{-1}$), $T_{\rm k}$ could be as low as $\sim 7.5$~K. Considering that $T_{\rm k}$ varies from 7.5 to 12 K, as 3$\sigma$, we adopt an uncertainty of $T_{\rm k}$ to be $12\pm1.5$ K ($\sim$12\%). Thus, the overall uncertainty of the HINSA column densities through error propagation should be 20\%. We note that the methodology we used to fit the HINSA signature is highly sensitive to the noise level of the spectra \citep{Krco2008}. There are two sources (G158.77-33.30 and G172.57-18.11) where we could not obtain reliable HINSA signatures, which are possibly limited by the current sensitivity of the GALFA-H\,{\sc i} survey. 

The uncertainties on CH column densities are given by the {\sc radex} model fits. As seen from Fig.~\ref{fig:ch_mcmc}(b), the column density of CH would be underestimated if the gas density were much higher. We believe that this uncertainty should not be significant since the deviation is within 50\% even if the gas density is above $10^6$~cm$^{-2}$. However, the average gas density at the current resolution should be no more than a few times $10^4$~cm$^{-3}$, otherwise we would obtain much higher values of $N_{\rm H_2}$. If we compare the derived column densities with the canonical assumption of $T_{\rm ex}=-15$~K, the column densities are only reduced by $\sim 17$\% on average. This is not a large deviation compared to the median uncertainties from our modeling ($\sim 13$\%). The observed line intensity ratios between CH's hyperfine splitting lines are close to the intrinsic line ratio, suggesting that the application of the canonical assumption of $T_{\rm ex} = -15$~K to cold dense cores is still robust in computing the CH column densities. 

\subsection{Excitation of CH}\label{sec:excitation}

The excitation temperature of CH ranges from $-32$ to $-5$~K, with a median value of $-6.5$~K (Table \ref{table:tab3}), showing the ubiquitous level inversion of the 3.3 GHz CH lines even in PGCCs. The values of $T_{\rm ex}$ are in agreement with the previous statistics from quasar ON-OFF observations that the distribution of $T_{\rm ex}$ mostly concentrates between $\sim -10$~K and $0$~K \citep{Tang2021}. In contrast to the observations in massive star-forming regions, we did not see any enhancements in the observed emission of the 3264 MHz satellite line and, as a result, do not infer very high values of $T_{\rm ex}$ close to zero. This is reasonable since there are no sources of strong FIR radiation inside or around these targets, and thus no FIR pumping mechanism would impact the level population, suggesting that the energy level inversion of CH in PGCCs is mainly due to the collisions with H$_2$ \citep{Rydbeck1976,Jacob2021}.

\subsection{Abundances}\label{sec:abundances}

\begin{figure}
\centering
\includegraphics[width=0.95\linewidth]{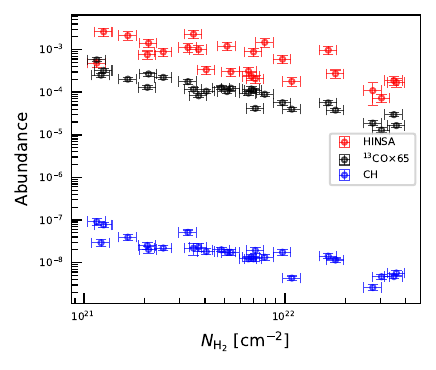}
\caption{The abundance of HINSA (red), CO (black), and CH (blue) as a function of $N_{\rm H_2}$. \label{fig:abundance}}
\end{figure}

The abundances of each species are calculated from the column densities ($N_{\rm col}$) divided by $N_{\rm H_2}$ ($f=N_{\rm col}/N_{\rm H_2}$). Figure \ref{fig:abundance} shows the comparison of abundances of HINSA, $^{13}$CO, and CH as a function of $N_{\rm H_2}$. All species show a decreasing trend as $N_{\rm H_2}$ increases, although for different reasons. The HINSA abundance $f_{\rm HINSA}$\footnote{Note that the definition of HINSA abundance here is different from the frequently used HINSA fraction in the literature, in which the latter is defined as $N_{\rm HINSA}/N_{\rm H} = f_{\rm HINSA}/2$.} is in the range of $(7.2\pm1.6)\times10^{-5}$ to $(2.6\pm0.6)\times10^{-3}$, with a mean value of $7.4\times10^{-4}$. This is slightly lower than previous HINSA surveys in nearby molecular clouds \citep[$\sim10^{-3}$, see][]{Li2003,Krco2008,Krco2010}. We attribute this to the better estimation of H$_2$ column density through {\it Herschel} and {\it Planck}, while the estimation from molecular lines such as CO isotopologues could be underestimated due to depletion (see below). Since the abundance of HINSA is proportional to the cosmic-ray ionization rate ($\zeta_2$, the ionization rate per H$_2$ molecule) and inversely proportional to $n_{\rm H_2}$ \citep{Goldsmith2005}, the decreasing trend mainly reflects the fact that $n_{\rm H_2}$ generally increases with $N_{\rm H_2}$, while $\zeta_2$ has the opposite trend \citep{Padovani2018a,Padovani2020,Padovani2024a}. 

The abundance of $^{13}$CO ranges from $(2.0\pm0.2)\times10^{-7}$ to $(8.9\pm1.0)\times10^{-6}$. If we assume an isotopic ratio of $^{12}$C/$^{13}$C equal to $66\pm 12$ \citep{Giannetti2014}, we find the abundance of CO to fall below $10^{-4}$ when $N_{\rm H_2}$ is above $10^{22}$~cm$^{-2}$. This is because CO freezes onto dust grains at high column densities, and the CO depletion factor increases with column density \citep{Caselli1999,Bacmann2002,Lippok2013}. However, the abundances of CO toward the two sources with the lowest values of $n_{\rm H_2}$ and $N_{\rm H_2}$ (G171.67-18.05 and G171.80-09.78) are greater than $2\times 10^{-4}$. This is unexpected since the abundance of CO is supposed to be low due to efficient photodissociation at $N_{\rm H_2} \sim 10^{21}$~cm$^{-2}$ \citep{Luo2023b}. Two possibilities can lead to this result; 1) the $^{12}$CO/$^{13}$CO isotopic ratio could be overestimated by a factor of few at such column densities, as it is known from both observations and chemical models \citep{Liszt2007,Szucs2014,Colzi2020,Sipila2023}; and 2) the values of $N_{\rm H_2}$ toward these sources could be underestimated due to the larger angular resolution of {\it Planck}. If the latter is the dominant reason, the column densities of H$_2$ toward G171.67-18.05 and G171.80-09.78 could be underestimated by factors of $\sim 2.2$ and 3.8, respectively. 

The abundance of CH is in the range of $(2.6\pm 0.4)\times10^{-9}$ to $(9.1\pm 1.7)\times10^{-8}$. These values are generally lower than the abundance in diffuse and translucent clouds, but are consistent with the measurements in high column density clouds \citep[$N_{\rm H_2} > 10^{21}$~cm$^{-2}$,][]{Qin2010}. Similar to $^{13}$CO, the abundance of CH is anomalously high toward the sources with the two lowest $N_{\rm H_2}$. 
If we use $^{13}$CO column densities to estimate H$_2$ column density (assuming CO/H$_2$ = $1.5\times10^{-4}$) toward G171.67-18.05 and G171.80-09.78, the abundances of CH toward both sources drop to $\sim 3\times10^{-8}$, consistent with the mean value at low column density \citep{Sheffer2008}. The decreasing trend of CH abundance is likely because CH is destroyed at higher column densities or its formation is suppressed (see Sect.~\ref{sec:evolution of ch}). 

\subsection{Estimates of $\zeta_2$ and cloud evolution}\label{sec:crir}

\begin{figure*}
\centering
\includegraphics[width=0.95\linewidth]{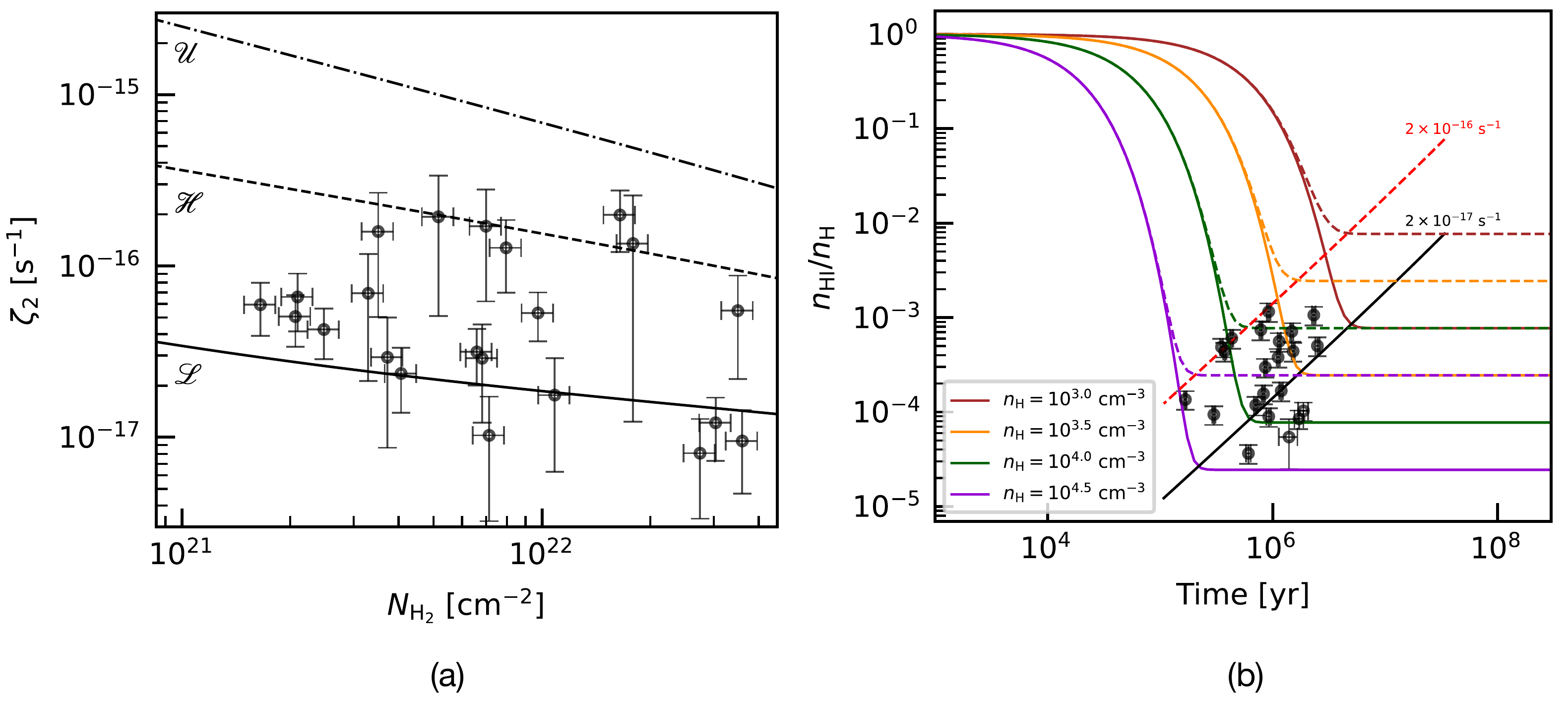}
\caption{(a): Values of $\zeta_2$ inferred from HINSA abundances. The black solid, dashed, and dash-dotted curves denote the theoretical CR attenuation models $\mathscr{L}$, $\mathscr{H}$, and $\mathscr{U}$ from \citet{Padovani2024b}. (b): Time dependence of atomic fraction at $n_{\rm H} = 10^{3.0}, 10^{3.5}, 10^{4.0}, {\rm and} \ 10^{4.5}$~cm$^{-3}$ with the assumption of $\zeta_2$ = $2\times10^{-17}$ (solid) and $2\times10^{-16}$ s$^{-1}$ (dashed), colored with orange, yellow, green, and purple. The black and red oblique lines represent two cutoffs at which solutions are close to chemical equilibrium under $\zeta_2$ = $2\times 10^{-17}$ and $2\times10^{-16}$~s$^{-1}$. \label{fig:hinsa_cr}}
\end{figure*}

Cosmic rays (CRs) are a primary driver of the chemical evolution in molecular clouds. However, constraining the values of $\zeta_2$ in molecular clouds using molecular line emission is difficult due to the complexity of the chemistry. The abundances of HINSA can provide independent constraints on $\zeta_2$ since the H\,{\sc i} inside the molecular cloud is a by-product of the dissociation of H$_2$ by CRs \citep{Goldsmith2005,Padovani2018b}. With the assumption of chemical equilibrium, $\zeta_2$ can be written as a function of HINSA abundance:
\begin{equation}
    \zeta_2 = \frac{1}{\eta}2k_{\rm H_2}n_{\rm H_2}f_{\rm HINSA},
\label{eq:zeta2}
\end{equation}
where $\eta = 1+0.7=1.7$ is a correction factor that takes into account 1) direct ionization of H$_2$ by CRs and 2) dissociation of H$_2$ primarily due to collisions with secondary CR electrons \citep{Padovani2018b}.
Here, $k_{\rm H_2}$ is the H$_2$ formation rate coefficient, for which we adopt a ``classical'' value of $3\times10^{-17}$~cm$^3$\,s$^{-1}$ \citep{Gry2002,LeBourlot2012,Bron2014,Wakelam2017}\footnote{Note that $k_{\rm H_2}$ could vary by an order of magnitude depending on environmental parameters \citep{Habart2004,Bron2014}. Given that the high value would only appear when the temperature is high (e.g., diffuse ISM where density is far below that hosted in PGCCs) in simulations \citep[e.g.,][]{Bron2014}, the current choice is still adequate. In contrast, a lower $k_{\rm H_2}$ adopted by many previous works would reduce the calculated $\zeta_2$.}. Since PGCCs are predominantly molecular, the H$_2$ volume density can be treated as $n_{\rm H_2} = n_{\rm H}/2$. We also ignore photodissociation by interstellar radiation field ($\zeta_{\rm pd}$) since this effect can be completely ignored even for the lowest column density targets in our sample \citep[$\zeta_{\rm pd} \ll 10^{-18}$~s$^{-1}$ at $N_{\rm H} > 4\times10^{21}$~cm$^{-2}$,][]{Padovani2018b}. The derived values of $\zeta_2$ range from $(8.1\pm4.7)\times10^{-18}$ to $(2.0\pm0.8)\times10^{-16}$ s$^{-1}$. Figure \ref{fig:hinsa_cr} (a) shows the derived $\zeta_2$ as a function of $N_{\rm H_2}$, with the CR attenuation models $\mathscr{L}$, $\mathscr{H}$, and $\mathscr{U}$ from \citet{Padovani2024b} overlaid for comparison. Although most of the values lie between model $\mathscr{L}$ and model $\mathscr{H}$, it is difficult to identify a trend.

Note that the above calculation of $\zeta_2$ only represents an upper limit if the cloud is far from chemical equilibrium \citep{Goldsmith2005,Goldsmith2007}. We follow the time-dependent equation for the atomic fraction ($n_{\rm HI}/n_{\rm H}$, here we assume HINSA has similar pathlength as H$_2$ so that $n_{\rm HI}/n_{\rm H} \approx N_{\rm HINSA}/N_{\rm H} = f_{\rm HINSA}/2$), as shown in \citet{Goldsmith2005}:
\begin{equation}
    \frac{n_{\rm HI}}{n_{\rm H}} = 1-\frac{2k_{\rm H_2}n_{\rm H}}{2k_{\rm H_2}n_{\rm H}+\eta\zeta_2}\left [1-e^{-t(2k_{\rm H_2}n_{\rm H}+\eta\zeta_2)}\right ].
    \label{eq:fch-t}
\end{equation}
We solve the equation for each target with $\eta\zeta_2 = 10^{-17}$~s$^{-1}$ (small enough to obtain a solution for the source with the lowest $\zeta_2$ value), given that $k_{\rm H_2}n_{\rm H}\gg\eta\zeta_2$ and the solved timescale would not largely impact by the exact value of $\zeta_2$. The resultant timescale for each PGCC is shown in Fig.~\ref{fig:hinsa_cr}(b). The evolution timescale ranges from $1.7\times10^5$ to $4.2\times10^6$ yr, with a mean value of $1.6\times10^6$ yr. 
If we adopt representative CR ionization rate of $\zeta_2 = 2\times10^{-17}$~s$^{-1}$ for model $\mathscr{L}$ and $\zeta_2 = 2\times10^{-16}$~s$^{-1}$ for model $\mathscr{H}$ at $N_{\rm H_2} \sim 10^{22}$~cm$^{-2}$, sources located around and below the black solid and red dashed lines are expected to achieve -- or approach to -- chemical equilibrium under the specified $\zeta_2$. 
The result suggests that when we consider model $\mathscr{H}$ as the representative CR attenuation model in PGCCs, almost all the samples should have reached chemical equilibrium. 

The fact that all the above factors would systematically overestimate $\zeta_2$ and that HINSA abundances provide stringent upper limits on $\zeta_2$ seem to favor CR attenuation model $\mathscr{L}$, in which the interstellar spectra of CR protons and electrons is extrapolated from {\it Voyager} \citep{Ivlev2015,Padovani2024b}. This aligns with the recent re-evaluation of $\zeta_2$ in diffuse clouds, where revised estimates exhibit closer agreement with model $\mathscr{L}$ \citep{Obolentseva2024,Neufeld2024}. 

We note that the exact volume density range traced by HINSA remains uncertain. We suspect that our adopted values of $n_{\rm H}$ are unlikely to be severely underestimated. The non-thermal velocity dispersion of HINSA is comparable to—or even lower than—that of $^{13}$CO (Fig.~\ref{fig:sigma_NT}), implying that most HINSA components arise from gas at higher densities than those typically traced by $^{13}$CO (i.e., above a few hundred cm$^{-3}$). \citet{Li2003} fitted the HINSA linewidths using polynomial functions and found that their non-thermal components generally lie between those of $^{13}$CO and C$^{18}$O, suggesting that HINSA predominantly traces intermediate-density gas (i.e., $10^3 \lesssim n_{\rm H} \lesssim 10^5$~cm$^{-3}$). From Eq.~\ref{eq:zeta2}, one can express: $N_{\rm HINSA} = \eta/(2k_{\rm H_2})\int \zeta_2dL$, where $L$ is the pathlength. Since high‐density cores (e.g., $n_{\rm H}>10^5$~cm$^{-3}$) exhibit much lower $\zeta_2$ values and occupy shorter path lengths than their surrounding envelopes, they are not expected to contribute substantially to the total $N_{\rm HINSA}$. For the $N_{\rm H_2}$ values in our sample, magnetohydrodynamical simulations (see Eq.~(5) in \citealt{Bisbas2023}) predict an average density of $n_{\rm H}\approx3\times10^2$ to $1.3\times10^4$~cm$^{-3}$, which is in good agreement with our adopted values. Consequently, we do not expect our estimates of $\zeta_2$ to be significantly underestimated. However, it remains possible that the HINSA column densities themselves are underestimated because of limited resolution and sensitivity. At this stage, we cannot yet rule out other CR attenuation models based on current data. Future high-sensitivity H\,{\sc i} and molecular line observations will be crucial to allow definitive conclusions. 

\subsection{The evolution of CH in PGCCs}\label{sec:evolution of ch}

In contrast to observations in diffuse clouds or massive star-forming regions, CH emission in PGCCs exhibits ubiquitous subsonic or transonic velocity dispersions, suggesting that the emission likely originates from the coherent dense gas structures (e.g., $n_{\rm H} \gtrsim 10^4$~cm$^{-3}$). 
In particular, approximately half of the subsonic components are observed in conjunction with supersonic components, potentially corresponding to the dense cores and their surrounding low-density components, as previously reported in TMC-1 \citep{Sakai2012}. In contrast, the remaining half of the sources exhibit only a single subsonic component. This discrepancy may indicate either that the observational sensitivity is insufficient to detect the weaker, broader linewidth components or that the ambient gas column density around dense cores is too low in these cases. 

Unlike most of the dense gas tracers (e.g., NH$_3$, N$_2$H$^+$) where molecular abundances accumulate with increasing column density \citep{Hotzel2004}\footnote{NH$_3$ and N$_2$H$^+$ are depleted at the center of prestellar cores \citep[e.g., if $N_{\rm H_2} > 2.6\times10^{22}$~cm$^{-2}$, $n_{\rm H_2} > 2\times10^5$~cm$^{-3}$,][]{Bergin2002,Pineda2022,Lin2023}}, the abundance of CH exhibits a monotonic decline with increasing column densities. This behavior can be attributed to two factors. First, efficient UV photon shielding with increasing column density facilitates rapid reactions between CH and abundant neutral species (e.g., atomic C and O), leading to the production of CO and more complex molecules at moderate densities \citep[$n_{\rm H} \sim 10^4$~cm$^{-3}$,][]{Van1988,Herbst1989}. Second, its formation is suppressed in high-density environments. One of the key CH formation channels that involves CH$^+$ is linked to non-equilibrium processes such as turbulent dissipation \citep{Federman1996, Godard2009}. However, in high column density environments, magnetohydrodynamic (MHD) turbulence cannot propagate efficiently into dense cores. This suppression of turbulent energy input disrupts the formation pathways of CH precursors, thereby reducing the abundance of CH in high-density regions.

\begin{figure}
\centering
\includegraphics[width=0.95\linewidth]{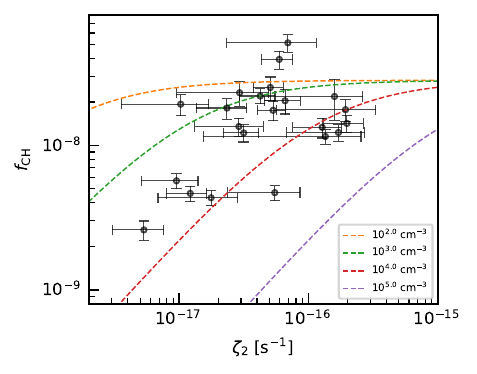}
\caption{The correlation between the abundance of CH and the values of $\zeta_2$ inferred from HINSA. The colored curves represent the predicted abundance of CH through Eq.~\ref{eq:fch} for different values of $n_{\rm H}$ as labelled. \label{fig:ch_cr}}
\end{figure}

Figure \ref{fig:ch_cr} shows a positive correlation between the abundance of CH and the $\zeta_2$ values inferred from HINSA, suggesting that CH formation is likely related to CR ionization. This is consistent with chemical simulations of regions with high column density, in which the abundance of CH increases with increasing $\zeta_2$ \citep{Luo2023a}. Considering the balance between CH formation and destruction, the steady-state abundance of CH can be written as (see Appendix \ref{sec:reactions} for more details):
\begin{equation}
    f_{\rm CH} = \frac{\alpha f_{\rm C^+}k_1}{f_{\rm HI}k_2+f_{\rm O}k_3},
    \label{eq:fch}
\end{equation}
where $\alpha = 0.3$ is the branching ratio for the dissociative recombination of CH$_3^+$ with electrons that leads to the formation of CH. 
All reaction rates ($k_1$ to $k_3$, listed in Table~\ref{table:rates}) are taken from the KInetic Database for Astrochemistry \citep[KIDA,][]{Wakelam2024} and calculated at $T_{\rm gas} = 12$~K. We replace $f_{\rm HINSA}$ using Eq.~\ref{eq:zeta2} and define $f_{\rm C^+}$ to be proportional to $\zeta_2/n_{\rm H_2}$. Fitting the observations with Eq.~\ref{eq:fch} yields constraints on the abundance of O and C$^+$ with respect to H$_2$, resulting in $f_{\rm O} = (4.8\pm0.8)\times10^{-4}$ and $f_{\rm C^+} = (7.4\pm0.7)\times10^{13}\zeta_2/n_{\rm H_2}$. The atomic oxygen abundance relative to total hydrogen is therefore $(2.4\pm0.4)\times10^{-4}$, which is in remarkable agreement with recent measurements in the ISM, $(2.51\pm0.69)\times10^{-4}$ \citep{Lis2023}. This suggests the potential to indirectly constrain the abundance of atomic oxygen. 

When $\zeta_2$ is low, atomic O dominates the destruction of CH, and $f_{\rm CH}$ scales linearly with $\zeta_2$. However, when atomic hydrogen (or C$^+$) become the dominant destruction partners, this linear relationship breaks down. The result suggests that atomic oxygen is unlikely to be heavily depleted in our sample. Otherwise, we would not expect to observe a positive correlation between CH and $\zeta_2$, as both formation and destruction of CH would then scale linearly with it, erasing the observed dependence. This is consistent with the result proposed by \citet{Caselli2002} that atomic oxygen should remain in the gas phase with an abundance in the order of $10^{-4}$ in dense core. 

\begin{figure}
\centering
\includegraphics[width=0.95\linewidth]{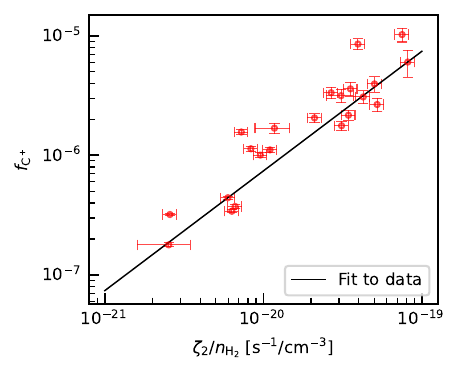}
\caption{The derived abundance of C$^+$ as a function of $\zeta_2/n_{\rm H_2}$. The black solid line denotes the fit to the data. \label{fig:fcp}}
\end{figure}

Figure \ref{fig:fcp} further illustrates how the derived $f_{\rm C^+}$ varies as a function of $\zeta_2/n_{\rm H_2}$. The abundance of C$^+$ in our sample spans from 10$^{-7}$ to 10$^{-5}$. The correlation shown in Fig.~\ref{fig:fcp} imposes an upper limit on the CH abundance under conditions where H dominate the destruction, yielding a maximum value of $\sim3\times 10^{-8}$. Furthermore, the correlation indicates that C$^+$ may serve as one of the main ions in dense clumps. In the case of a dense core such as L1544 with $\zeta_2/n_{\rm H} \sim {\rm a\ few} 10^{-23}$~cm$^3$~s$^{-1}$, the abundance of C$^+$ (a few $\sim10^{-9}$) remains comparable to — or even exceeds — that of commonly assumed dominant ions such as H$_3^+$, HCO$^+$, N$_2$H$^+$, etc.

While we demonstrate that the formation pathway of CH through the reaction of C$^+$ with H$_2$ surpasses that of C with H$_3^+$, it is known that non-equilibrium chemistry plays a crucial role in the formation of CH$^+$ \citep{Federman1996, Godard2009, Godard2023}. Furthermore, turbulent diffusive transport can efficiently bring different molecules and ions into denser regions, significantly altering their abundances in higher column density environments \citep{Xie1995,Scalo2004}. An enhancement of CH$^+$ abundance in the order of 10$^{-12}$ in our targets can contribute to a non-negligible effect on the formation of CH. Therefore, in regions of lower column density where MHD turbulence can effectively penetrate, the CH abundance estimated using Eq.~\ref{eq:fch} could be underestimated. 
In summary, the decreasing abundance of CH shown in Fig.~\ref{fig:abundance} is mainly due to a decrease in $\zeta_2$ that reduces the gas-phase abundance of C$^+$, which is critical for CH formation; on the other hand, the lower ionization fraction suppresses the MHD turbulence generated by streaming cosmic rays and its dissipation into smaller scales \citep{Ivlev2018,Padovani2020}, thereby suppressing the abundance of key CH precursors like CH$^+$. 

\section{Conclusion}\label{sec:conclusion}

We have performed CH single-pointing observations toward 27 PGCCs with the Arecibo 305m telescope. We analyzed the CH emission together with archival $^{13}$CO (1--0) and H\,{\sc i} 21 cm emission line data, and calculated the non-thermal velocity dispersions and abundances. The main conclusions are as follows:
\begin{enumerate}

\item The velocity dispersions of both HINSA and CH are generally lower than those of $^{13}$CO. In particular, half of the CH components exhibit subsonic or transonic non-thermal velocity dispersions; in contrast, almost all $^{13}$CO components are supersonic. This result suggests that, for cold, dense cores, the narrow-line CH emission predominantly traces gas in regions of lower turbulence and higher density compared to $^{13}$CO.

\item The excitation temperature of CH towards PGCCs from {\sc radex} modeling has a median value of $-6.5$~K, in agreement with previous statistical results. The deviation of $T_{\rm ex}$ from the canonical assumption $T_{\rm ex} = -15$~K introduces an uncertainty less than 17\%, suggesting that the canonical assumption is still robust in cold, dense clumps.

\item The abundance of $^{13}$CO decreases with increasing $N_{\rm H_2}$. Assuming $^{12}{\rm C}/^{13}{\rm C} = 66 \pm 12$, the abundance of CO declines below the canonical value ($\sim 10^{-4}$) when $N_{\rm H_2} > 10^{22}$~cm$^{-2}$, indicating that CO freeze-out onto grain surfaces becomes efficient enough to significantly reduce the gas-phase CO reservoir in the densest cloud interiors. 

\item The derived values of $\zeta_2$ from HINSA under the assumption of chemical equilibrium span from $(8.1\pm4.7)\times 10^{-18}$ to $(2.0\pm0.8)\times10^{-16}$~s$^{-1}$. Since these estimates represent upper limits, they suggest that the ionization rates in PGCCs are consistent with the theoretically predicted CR attenuation model $\mathscr{L}$, in agreement with the recent re-evaluation of $\zeta_2$ in diffuse clouds.

\item We fit the abundance of CH with an analytic formula based on chemical equilibrium, yielding constraints on the gas phase abundance of O and C$^+$. The positive correlation between CH and $\zeta_2$ is due to the fact that atomic oxygen is the main destroyer of CH in low ionization and high column density environments. The decline in abundance of CH with $N_{\rm H_2}$ may be related to both the decrease in abundance of C$^+$ and the suppression of abundance of CH$^+$. 

\end{enumerate}

Future high-sensitivity, high-angular-resolution observations with large facilities such as FAST, ngVLA, and SKA will play a pivotal role in establishing stringent constraints on CR attenuation models and determining the specific density regimes and spatial origins from which the majority of HINSA and CH signatures originate. These observations are expected to reveal a pronounced decline in HINSA and CH abundance distributions toward the central regions of prestellar cores. Furthermore, the characteristic depletion scale of these species — the spatial extent over which their abundances diminish significantly — may coincide with the transition zone where MHD turbulence ceases to propagate effectively. This transition zone, known as the coherent structure and first identified by \citet{Goodman1998} using NH$_3$ observations, likely marks the boundary of the dynamically quiescent region where prestellar core formation occurs.


\begin{acknowledgements}
We thank the scientific staff and telescope operators at Arecibo Observatory (AO) and Purple Mountain Observatory (PMO), particularly Wenting Xu, for their advice and assistance with observations and data reduction.
This publication makes use of data products from the 13.7m telescope of the Qinghai observation station of the Purple Mountain Observatory and the millimeter wave radio astronomy database. 
This publication utilizes data from Galactic ALFA HI (GALFA HI) survey data set obtained with the Arecibo L-band Feed Array (ALFA) on the Arecibo 305m telescope. The Arecibo Observatory is operated by SRI International under a cooperative agreement with the National Science Foundation (AST-1100968), and in alliance with Ana G. Méndez-Universidad Metropolitana, and the Universities Space Research Association. The GALFA HI surveys have been funded by the NSF through grants to Columbia University, the University of Wisconsin, and the University of California.
This research has made use of data from the Herschel Gould Belt survey (HGBS) project (http://gouldbelt-herschel.cea.fr). The HGBS is a Herschel Key Programme jointly carried out by SPIRE Specialist Astronomy Group 3 (SAG 3), scientists of several institutes in the PACS Consortium (CEA Saclay, INAF-IFSI Rome and INAF-Arcetri, KU Leuven, MPIA Heidelberg), and scientists of the Herschel Science Center (HSC).
MP acknowledges the INAF-Minigrant 2024 ENERGIA (``ExploriNg low-Energy cosmic Rays throuGh theoretical InvestigAtions at INAF'');
DG acknowledges the INAF-Minigrant 2023 PACIFISM (``PArtiCles, Ionization and Fields in the InterStellar Medium'').
\end{acknowledgements}

%
%

\bibliographystyle{aa}
\bibliography{reference} 



\begin{appendix} 


\section{Source properties}\label{sec:sources}

\begin{table*}
\caption{Summary of source properties.}           
\label{table:tab1}      
\centering  
\begin{tabular}{lcccc}
\hline\hline
source name & RA(hms), Dec(dms) & $T_{\rm dust}$/K & $\log_{10}$($\frac{n_{\rm H}}{{\rm cm}^{-3}}$) & $\log_{10}$($\frac{N_{\rm H_2}}{{\rm cm}^{-2}}$) \\
\hline
G158.20-20.28 & 03:29:19,+31:33:55 & 12.83 & $3.84 \pm 0.23$ & $21.83 \pm 0.04$ \\
G158.22-20.14 & 03:29:48,+31:39:49 & 13.36 & -- & $21.73 \pm 0.04$ \\
G158.40-21.86 & 03:25:36,+30:11:30 & 12.75 & $4.04 \pm 0.26$ & $21.84 \pm 0.04$ \\
G158.77-33.30* & 02:57:33,+20:38:31 & 12.46 & $3.43 \pm 0.24$ & $21.68 \pm 0.04$ \\
G158.86-21.60 & 03:27:54,+30:08:34 & 12.38 & $3.96 \pm 0.30$ & $21.71 \pm 0.04$ \\
G158.88-34.18 & 02:55:48,+19:51:51 & 12.79 & $3.45 \pm 0.28$ & $21.85 \pm 0.04$ \\
G166.99-15.34 & 04:13:42,+29:44:26 & 12.92 & $3.22 \pm 0.29$ & $21.57 \pm 0.04$ \\
G167.23-15.32 & 04:14:31,+29:35:16 & 12.20 & $3.70 \pm 0.10$ & $21.99 \pm 0.04$ \\
G168.00-15.69 & 04:15:40,+28:48:01 & 12.25 & $3.69 \pm 0.17$ & $21.90 \pm 0.04$ \\
G168.13-16.39 & 04:13:48,+28:13:22 & 11.63 & $4.45 \pm 0.38$ & $22.25 \pm 0.04$ \\
G168.72-15.48 & 04:18:35,+28:26:35 & 11.72 & $4.06 \pm 0.14$ & $22.22 \pm 0.04$ \\
G169.82-19.39 & 04:09:12,+24:58:39 & 12.07 & $3.75 \pm 0.26$ & $22.04 \pm 0.04$ \\
G169.84-07.61* & 04:49:23,+32:49:47 & 12.57 & $3.58 \pm 0.11$ & $21.32 \pm 0.04$ \\
G170.77-08.51* & 04:48:49,+31:33:00 & 12.95 & $3.20 \pm 0.11$ & $21.22 \pm 0.04$ \\
G170.88-10.92* & 04:40:33,+29:55:43 & 11.68 & $3.60 \pm 0.15$ & $21.61 \pm 0.04$ \\
G171.14-17.58 & 04:18:48,+25:18:56 & 11.13 & $3.98 \pm 0.14$ & $22.48 \pm 0.04$ \\
G171.67-18.05*$\dagger$ & 04:18:47,+24:37:40 & 12.50 & $3.15 \pm 0.12$ & $21.10 \pm 0.04$ \\
G171.80-09.78*$\dagger$ & 04:47:16,+29:57:34 & 12.77 & $3.06 \pm 0.19$ & $21.06 \pm 0.04$ \\
G171.84-05.22 & 05:04:09,+32:44:35 & 10.40 & $3.62 \pm 0.10$ & $22.44 \pm 0.04$ \\
G172.06-15.21 & 04:29:16,+26:14:52 & 12.43 & $3.76 \pm 0.12$ & $21.82 \pm 0.04$ \\
G172.57-18.11* & 04:21:07,+23:57:18 & 13.36 & -- & $21.08 \pm 0.04$ \\
G172.92-16.74 & 04:26:35,+24:36:59 & 11.14 & $4.22 \pm 0.24$ & $22.54 \pm 0.04$ \\
G173.07-16.52* & 04:27:43,+24:38:49 & 11.94 & $3.59 \pm 0.28$ & $21.55 \pm 0.04$ \\
G173.07-17.89* & 04:23:14,+23:44:29 & 12.16 & $3.44 \pm 0.10$ & $21.39 \pm 0.04$ \\
G173.89-17.64* & 04:26:18,+23:19:47 & 12.48 & $3.42 \pm 0.13$ & $21.32 \pm 0.04$ \\
G173.91-16.25* & 04:30:56,+24:13:15 & 11.68 & $3.54 \pm 0.28$ & $21.52 \pm 0.04$ \\
G175.58-16.60 & 04:34:17,+22:46:22 & 10.07 & $3.50 \pm 0.20$ & $22.56 \pm 0.04$ \\
\hline
\end{tabular}
\tablefoot{From left to right the columns are: source name, coordinate, dust temperature, total H volume density, and H$_2$ column density. ``*'' indicates that the column density of H$_2$ is obtained from {\it Planck}. ``$\dagger$'' indicates the two sources where the column density may be underestimated, and they are not involved in the analysis and discussion in the main text.}
\end{table*}

\section{Gaussian fitting results of $^{13}$CO}\label{sec:gsfit_13co}

\begin{table*}
\caption{Source name, velocity, linewidth, brightness temperature, column density, and optical depth of $^{13}$CO.}      
\label{table:tab2}      
\centering 
\begin{tabular}{lccccc}
\hline\hline
source name & $V_{\rm lsr}$ & $\Delta V$ & $T_{\rm mb}$ & $\log_{10}$($N_{\rm ^{13}CO}$/cm$^{-2}$) & $\tau_{\rm ^{13}CO}$ \\
 & (km\,s$^{-1}$) & (km\,s$^{-1}$) & (K) &  &  \\
\hline
G158.20-20.28 & $1.72 \pm 0.04$ & $1.21 \pm 0.08$ & $0.68 \pm 0.04$ & $14.99 \pm 0.04$ & $0.07 \pm 0.00$ \\
 & $6.59 \pm 0.06$ & $1.55 \pm 0.11$ & $1.68 \pm 0.05$ & $15.51 \pm 0.03$ & $0.20 \pm 0.01$ \\
 & $7.60 \pm 0.01$ & $0.76 \pm 0.02$ & $5.02 \pm 0.20$ & $15.79 \pm 0.03$ & $0.75 \pm 0.05$ \\
 & $8.61 \pm 0.06$ & $1.27 \pm 0.11$ & $1.35 \pm 0.05$ & $15.32 \pm 0.04$ & $0.15 \pm 0.01$ \\
G158.22-20.14 & $8.18 \pm 0.01$ & $1.03 \pm 0.01$ & $5.79 \pm 0.06$ & $16.01 \pm 0.01$ & $0.86 \pm 0.02$ \\
G158.40-21.86 & $4.37 \pm 0.01$ & $1.22 \pm 0.02$ & $5.66 \pm 0.06$ & $16.08 \pm 0.01$ & $0.92 \pm 0.02$ \\
G158.77-33.30 & $-2.65 \pm 0.01$ & $1.47 \pm 0.01$ & $4.02 \pm 0.03$ & $15.94 \pm 0.01$ & $0.58 \pm 0.01$ \\
 & $1.11 \pm 0.03$ & $1.12 \pm 0.07$ & $0.56 \pm 0.03$ & $14.86 \pm 0.04$ & $0.06 \pm 0.00$ \\
G158.86-21.60 & $5.55 \pm 0.00$ & $2.16 \pm 0.01$ & $2.83 \pm 0.01$ & $15.92 \pm 0.00$ & $0.38 \pm 0.00$ \\
G158.88-34.18 & $-5.39 \pm 0.01$ & $0.69 \pm 0.03$ & $0.98 \pm 0.04$ & $14.91 \pm 0.03$ & $0.11 \pm 0.01$ \\
 & $-4.37 \pm 0.00$ & $0.79 \pm 0.02$ & $2.02 \pm 0.06$ & $15.31 \pm 0.02$ & $0.24 \pm 0.01$ \\
 & $-4.21 \pm 0.06$ & $2.33 \pm 0.08$ & $0.64 \pm 0.06$ & $15.24 \pm 0.04$ & $0.07 \pm 0.01$ \\
G166.99-15.34 & $6.45 \pm 0.02$ & $1.33 \pm 0.04$ & $2.66 \pm 0.06$ & $15.68 \pm 0.02$ & $0.33 \pm 0.01$ \\
G167.23-15.32 & $4.50 \pm 0.06$ & $0.99 \pm 0.15$ & $0.71 \pm 0.08$ & $14.91 \pm 0.08$ & $0.08 \pm 0.01$ \\
 & $6.31 \pm 0.01$ & $1.30 \pm 0.03$ & $3.94 \pm 0.07$ & $15.88 \pm 0.02$ & $0.59 \pm 0.01$ \\
 & $8.01 \pm 0.09$ & $0.45 \pm 0.19$ & $0.31 \pm 0.12$ & $14.19 \pm 0.25$ & $0.04 \pm 0.01$ \\
G168.00-15.69 & $7.60 \pm 0.01$ & $0.88 \pm 0.02$ & $6.36 \pm 0.12$ & $16.04 \pm 0.02$ & $1.25 \pm 0.05$ \\
G168.13-16.39 & $6.02 \pm 0.01$ & $0.61 \pm 0.03$ & $4.16 \pm 0.12$ & $15.58 \pm 0.03$ & $0.70 \pm 0.03$ \\
 & $6.88 \pm 0.01$ & $0.78 \pm 0.03$ & $5.02 \pm 0.10$ & $15.82 \pm 0.02$ & $0.93 \pm 0.03$ \\
G168.72-15.48 & $5.49 \pm 0.02$ & $0.57 \pm 0.05$ & $2.05 \pm 0.16$ & $15.16 \pm 0.05$ & $0.28 \pm 0.02$ \\
 & $7.17 \pm 0.01$ & $0.90 \pm 0.03$ & $6.41 \pm 0.13$ & $16.08 \pm 0.02$ & $1.45 \pm 0.06$ \\
 & $8.38 \pm 0.10$ & $0.88 \pm 0.24$ & $0.77 \pm 0.13$ & $14.89 \pm 0.14$ & $0.10 \pm 0.02$ \\
G169.82-19.39 & $7.97 \pm 0.01$ & $0.84 \pm 0.02$ & $4.90 \pm 0.13$ & $15.82 \pm 0.02$ & $0.83 \pm 0.03$ \\
G169.84-07.61 & $6.29 \pm 0.00$ & $0.67 \pm 0.01$ & $4.15 \pm 0.03$ & $15.62 \pm 0.01$ & $0.60 \pm 0.01$ \\
G170.77-08.51 & $5.98 \pm 0.03$ & $0.91 \pm 0.06$ & $0.90 \pm 0.02$ & $14.99 \pm 0.03$ & $0.10 \pm 0.00$ \\
 & $6.70 \pm 0.00$ & $0.55 \pm 0.01$ & $4.17 \pm 0.06$ & $15.54 \pm 0.01$ & $0.57 \pm 0.01$ \\
 & $7.67 \pm 0.01$ & $0.57 \pm 0.02$ & $1.06 \pm 0.03$ & $14.86 \pm 0.02$ & $0.12 \pm 0.00$ \\
G170.88-10.92 & $2.73 \pm 0.11$ & $1.78 \pm 0.25$ & $0.58 \pm 0.07$ & $15.07 \pm 0.08$ & $0.07 \pm 0.01$ \\
 & $5.97 \pm 0.01$ & $0.66 \pm 0.02$ & $4.91 \pm 0.12$ & $15.73 \pm 0.02$ & $0.89 \pm 0.03$ \\
G171.14-17.58 & $3.30 \pm 0.01$ & $0.51 \pm 0.04$ & $2.42 \pm 0.15$ & $15.20 \pm 0.05$ & $0.37 \pm 0.03$ \\
 & $7.29 \pm 0.01$ & $0.81 \pm 0.03$ & $3.75 \pm 0.11$ & $15.65 \pm 0.02$ & $0.65 \pm 0.03$ \\
G171.67-18.05 & $3.78 \pm 0.01$ & $1.04 \pm 0.03$ & $4.05 \pm 0.09$ & $15.79 \pm 0.02$ & $0.59 \pm 0.02$ \\
G171.80-09.78 & $5.72 \pm 0.00$ & $0.64 \pm 0.01$ & $7.02 \pm 0.10$ & $15.97 \pm 0.01$ & $1.37 \pm 0.04$ \\
 & $6.86 \pm 0.03$ & $0.66 \pm 0.07$ & $1.17 \pm 0.09$ & $14.97 \pm 0.06$ & $0.13 \pm 0.01$ \\
G171.84-05.22 & $6.89 \pm 0.01$ & $0.52 \pm 0.02$ & $6.15 \pm 0.17$ & $15.90 \pm 0.04$ & $2.03 \pm 0.18$ \\
G172.06-15.21 & $6.70 \pm 0.01$ & $1.03 \pm 0.01$ & $5.49 \pm 0.07$ & $15.99 \pm 0.01$ & $0.93 \pm 0.02$ \\
G172.57-18.11 & $2.72 \pm 0.03$ & $1.28 \pm 0.06$ & $1.35 \pm 0.06$ & $15.33 \pm 0.03$ & $0.14 \pm 0.01$ \\
 & $4.85 \pm 0.01$ & $0.59 \pm 0.02$ & $3.13 \pm 0.09$ & $15.41 \pm 0.02$ & $0.38 \pm 0.01$ \\
G172.92-16.74 & $4.98 \pm 0.01$ & $0.73 \pm 0.02$ & $5.50 \pm 0.14$ & $15.87 \pm 0.02$ & $1.22 \pm 0.06$ \\
 & $6.75 \pm 0.01$ & $0.84 \pm 0.02$ & $5.46 \pm 0.13$ & $15.93 \pm 0.02$ & $1.20 \pm 0.05$ \\
G173.07-16.52 & $6.57 \pm 0.01$ & $0.99 \pm 0.03$ & $4.22 \pm 0.11$ & $15.80 \pm 0.02$ & $0.68 \pm 0.02$ \\
G173.07-17.89 & $3.92 \pm 0.01$ & $1.27 \pm 0.03$ & $4.37 \pm 0.09$ & $15.93 \pm 0.02$ & $0.68 \pm 0.02$ \\
G173.89-17.64 & $5.12 \pm 0.01$ & $0.84 \pm 0.02$ & $5.79 \pm 0.11$ & $15.94 \pm 0.02$ & $1.00 \pm 0.03$ \\
G173.91-16.25 & $6.12 \pm 0.01$ & $0.72 \pm 0.03$ & $6.20 \pm 0.22$ & $15.95 \pm 0.04$ & $1.36 \pm 0.10$ \\
G175.58-16.60 & $5.69 \pm 0.02$ & $1.12 \pm 0.04$ & $4.62 \pm 0.14$ & $15.96 \pm 0.03$ & $1.15 \pm 0.07$ \\
\hline
\end{tabular}
\end{table*}

\section{Gaussian fitting results of CH}\label{sec:gsfit_ch}

\begin{table*}
\caption{Source name, velocity, linewidth, brightness temperature, column density, and excitation temperature of CH 3335~MHz.}           
\label{table:tab3}      
\centering  
\begin{tabular}{lccccc}
\hline\hline
source name & $V_{\rm lsr}$ & $\Delta V$ & $T_{\rm mb}$ & $\log_{10}$($N_{\rm CH}$/cm$^{-2}$) & $T_{\rm ex}$ \\
 & (km\,s$^{-1}$) & (km\,s$^{-1}$) & (K) &  & (K) \\
\hline
G158.20-20.28 & $8.00 \pm 0.11$ & $2.91 \pm 0.26$ & $0.15 \pm 0.01$ & $13.96 \pm 0.04$ & $-5.63$ \\
G158.22-20.14 & $6.95 \pm 0.25$ & $1.40 \pm 0.53$ & $0.10 \pm 0.02$ & $13.45 \pm 0.07$ & $-5.04$ \\
 & $8.46 \pm 0.08$ & $1.17 \pm 0.15$ & $0.27 \pm 0.02$ & $13.82 \pm 0.03$ & $-4.98$ \\
G158.40-21.86 & $4.31 \pm 0.02$ & $0.91 \pm 0.05$ & $0.42 \pm 0.02$ & $13.93 \pm 0.04$ & $-6.10$ \\
G158.77-33.30 & $-2.44 \pm 0.03$ & $1.22 \pm 0.07$ & $0.38 \pm 0.02$ & $13.98 \pm 0.02$ & $-5.95$ \\
G158.86-21.60 & $5.27 \pm 0.08$ & $2.12 \pm 0.18$ & $0.19 \pm 0.01$ & $13.96 \pm 0.06$ & $-7.20$ \\
G158.88-34.18 & $-4.95 \pm 0.06$ & $1.27 \pm 0.14$ & $0.26 \pm 0.02$ & $13.99 \pm 0.06$ & $-12.08$ \\
 & $-3.24 \pm 0.12$ & $1.32 \pm 0.29$ & $0.14 \pm 0.02$ & $13.61 \pm 0.05$ & $-5.66$ \\
G166.99-15.34 & $7.01 \pm 0.07$ & $1.18 \pm 0.15$ & $0.30 \pm 0.02$ & $13.85 \pm 0.06$ & $-6.16$ \\
 & $8.10 \pm 0.16$ & $0.77 \pm 0.32$ & $0.09 \pm 0.03$ & $13.17 \pm 0.14$ & $-5.50$ \\
G167.23-15.32 & $6.11 \pm 0.10$ & $0.64 \pm 0.28$ & $0.13 \pm 0.04$ & $13.34 \pm 0.10$ & $-6.48$ \\
 & $6.82 \pm 0.02$ & $0.52 \pm 0.05$ & $0.58 \pm 0.04$ & $13.78 \pm 0.03$ & $-6.33$ \\
 & $7.45 \pm 0.22$ & $2.40 \pm 0.33$ & $0.17 \pm 0.02$ & $13.94 \pm 0.05$ & $-6.58$ \\
G168.00-15.69 & $7.81 \pm 0.01$ & $0.34 \pm 0.04$ & $0.45 \pm 0.05$ & $13.53 \pm 0.04$ & $-6.30$ \\
 & $8.07 \pm 0.05$ & $1.20 \pm 0.10$ & $0.28 \pm 0.04$ & $13.86 \pm 0.06$ & $-6.51$ \\
G168.13-16.39 & $6.53 \pm 0.00$ & $0.37 \pm 0.01$ & $1.61 \pm 0.03$ & $14.16 \pm 0.01$ & $-13.60$ \\
 & $7.09 \pm 0.02$ & $0.59 \pm 0.05$ & $0.46 \pm 0.02$ & $13.78 \pm 0.05$ & $-10.08$ \\
G168.72-15.48 & $7.20 \pm 0.15$ & $3.60 \pm 0.40$ & $0.13 \pm 0.02$ & $14.02 \pm 0.05$ & $-7.40$ \\
 & $7.47 \pm 0.01$ & $0.44 \pm 0.01$ & $1.29 \pm 0.03$ & $14.11 \pm 0.02$ & $-10.08$ \\
G169.82-19.39 & $8.27 \pm 0.02$ & $0.57 \pm 0.05$ & $0.37 \pm 0.03$ & $13.67 \pm 0.03$ & $-6.57$ \\
G169.84-07.61 & $6.49 \pm 0.03$ & $0.74 \pm 0.06$ & $0.31 \pm 0.02$ & $13.72 \pm 0.07$ & $-7.71$ \\
G170.77-08.51 & $6.13 \pm 0.08$ & $0.87 \pm 0.20$ & $0.16 \pm 0.02$ & $13.52 \pm 0.05$ & $-5.46$ \\
 & $6.96 \pm 0.03$ & $0.49 \pm 0.06$ & $0.35 \pm 0.03$ & $13.51 \pm 0.04$ & $-5.40$ \\
G170.88-10.92 & $6.11 \pm 0.02$ & $0.58 \pm 0.04$ & $0.46 \pm 0.03$ & $13.87 \pm 0.06$ & $-17.25$ \\
G171.14-17.58 & $3.60 \pm 0.03$ & $0.63 \pm 0.07$ & $0.24 \pm 0.02$ & $13.58 \pm 0.04$ & $-9.03$ \\
 & $7.60 \pm 0.01$ & $0.42 \pm 0.01$ & $1.05 \pm 0.03$ & $14.01 \pm 0.03$ & $-11.28$ \\
G171.67-18.05 & $3.84 \pm 0.04$ & $1.14 \pm 0.09$ & $0.26 \pm 0.02$ & $13.77 \pm 0.03$ & $-5.93$ \\
 & $7.42 \pm 0.09$ & $1.50 \pm 0.20$ & $0.13 \pm 0.02$ & $13.56 \pm 0.06$ & $-6.03$ \\
G171.80-09.78 & $6.01 \pm 0.04$ & $0.53 \pm 0.11$ & $0.28 \pm 0.03$ & $13.47 \pm 0.04$ & $-5.62$ \\
 & $6.45 \pm 0.05$ & $0.22 \pm 0.18$ & $0.13 \pm 0.05$ & $12.61 \pm 0.32$ & $-5.72$ \\
 & $7.07 \pm 0.03$ & $0.75 \pm 0.07$ & $0.35 \pm 0.02$ & $13.86 \pm 0.06$ & $-12.07$ \\
G171.84-05.22 & $7.14 \pm 0.01$ & $0.36 \pm 0.02$ & $0.70 \pm 0.03$ & $13.85 \pm 0.05$ & $-32.66$ \\
G172.06-15.21 & $6.95 \pm 0.02$ & $1.00 \pm 0.06$ & $0.38 \pm 0.02$ & $13.91 \pm 0.04$ & $-6.45$ \\
G172.57-18.11 & $2.85 \pm 0.06$ & $0.71 \pm 0.15$ & $0.13 \pm 0.02$ & $13.21 \pm 0.08$ & $-5.03$ \\
 & $5.06 \pm 0.04$ & $0.57 \pm 0.09$ & $0.20 \pm 0.03$ & $13.29 \pm 0.07$ & $-5.05$ \\
G172.92-16.74 & $5.17 \pm 0.05$ & $0.66 \pm 0.12$ & $0.16 \pm 0.03$ & $13.45 \pm 0.06$ & $-8.88$ \\
 & $6.93 \pm 0.01$ & $0.54 \pm 0.02$ & $1.03 \pm 0.03$ & $14.14 \pm 0.02$ & $-12.49$ \\
G173.07-16.52 & $6.65 \pm 0.17$ & $1.16 \pm 0.22$ & $0.15 \pm 0.06$ & $13.60 \pm 0.17$ & $-6.96$ \\
 & $6.95 \pm 0.03$ & $0.49 \pm 0.08$ & $0.37 \pm 0.07$ & $13.56 \pm 0.08$ & $-7.00$ \\
G173.07-17.89 & $4.32 \pm 0.03$ & $0.74 \pm 0.06$ & $0.31 \pm 0.02$ & $13.74 \pm 0.03$ & $-6.46$ \\
G173.89-17.64 & $5.29 \pm 0.02$ & $0.43 \pm 0.04$ & $0.41 \pm 0.03$ & $13.63 \pm 0.07$ & $-9.48$ \\
G173.91-16.25 & $6.04 \pm 0.05$ & $1.36 \pm 0.09$ & $0.30 \pm 0.02$ & $13.99 \pm 0.05$ & $-7.77$ \\
 & $6.41 \pm 0.00$ & $0.26 \pm 0.01$ & $1.32 \pm 0.04$ & $13.86 \pm 0.03$ & $-9.13$ \\
G175.58-16.60 & $5.82 \pm 0.02$ & $0.82 \pm 0.04$ & $0.76 \pm 0.02$ & $14.24 \pm 0.03$ & $-29.62$ \\
 & $6.60 \pm 0.03$ & $0.44 \pm 0.08$ & $0.25 \pm 0.03$ & $13.46 \pm 0.05$ & $-15.52$ \\
\hline
\end{tabular}
\end{table*}

\section{An example of HINSA fitting toward G168.13-16.39}\label{sec:mcmc_hinsa}

\begin{figure}
\centering
\includegraphics[width=0.95\linewidth]{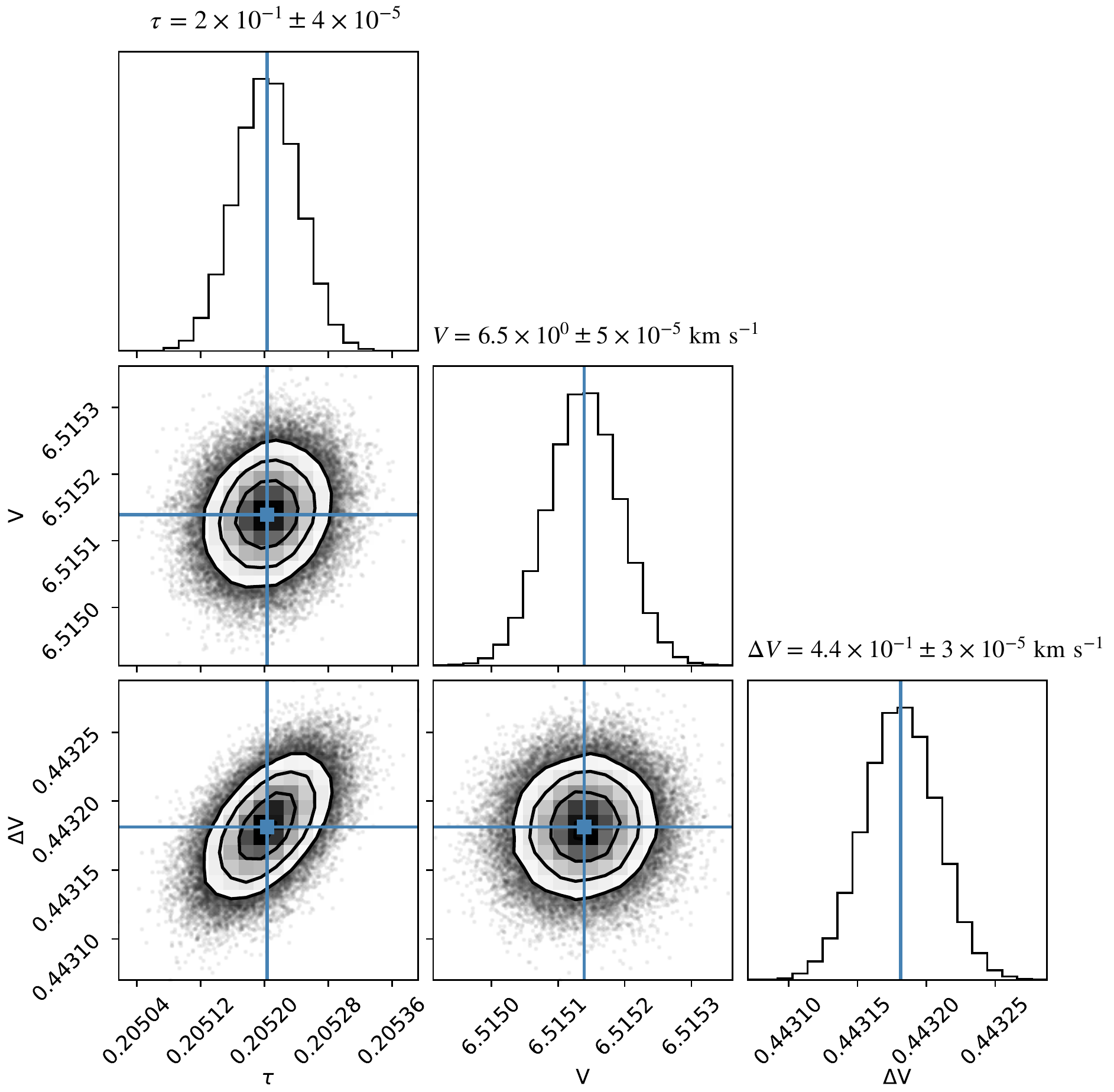}
\caption{The posterior probability distribution of $\tau$, $V_{\rm lsr}$, and $\Delta V$ of HINSA toward G168.13-16.39. \label{fig:hinsa_fit}}
\end{figure}

\section{HINSA fitting results}\label{sec:gsfit_hinsa}


\begin{table*}
\caption{Source name, velocity, linewidth, optical depth, column density, and $\zeta_2$ derived from HINSA fitting.}           
\label{table:tab4}      
\centering  
\begin{tabular}{lccccc}
\hline\hline
source name & $V_{\rm lsr}$ & $\Delta V$ & $\tau$ & $\log_{10}$($\frac{N_{\rm HINSA}}{\rm cm^{-2}}$) & $\log_{10}$($\frac{\zeta_2}{\rm s^{-1}}$) \\
 & (km\,s$^{-1}$) & (km\,s$^{-1}$) &  &   \\
\hline
G158.20-20.28 & 7.70 & 1.05 & 0.06 & $18.21 \pm 0.09$ & $-16.54 \pm 0.25$ \\
G158.22-20.14 & 8.15 & 1.01 & 0.06 & $18.21 \pm 0.09$ & -- \\
G158.40-21.86 & 4.06 & 1.44 & 0.17 & $18.79 \pm 0.09$ & $-15.77 \pm 0.28$ \\
G158.77-33.30 & -- & -- & -- & -- & -- \\
G158.86-21.60 & 4.96 & 1.67 & 0.15 & $18.79 \pm 0.09$ & $-15.71 \pm 0.32$ \\
G158.88-34.18 & -5.12 & 1.27 & 0.05 & $18.17 \pm 0.09$ & $-16.99 \pm 0.30$ \\
G166.99-15.34 & 6.74 & 1.26 & 0.12 & $18.57 \pm 0.09$ & $-16.53 \pm 0.31$ \\
G167.23-15.32 & 6.91 & 1.49 & 0.16 & $18.76 \pm 0.09$ & $-16.27 \pm 0.14$ \\
G168.00-15.69 & 7.83 & 1.19 & 0.42 & $19.07 \pm 0.09$ & $-15.89 \pm 0.20$ \\
G168.13-16.39 & 6.51 & 1.04 & 0.21 & $18.69 \pm 0.09$ & $-15.87 \pm 0.39$ \\
G168.72-15.48 & 7.34 & 1.25 & 0.56 & $19.21 \pm 0.09$ & $-15.70 \pm 0.17$ \\
G169.82-19.39 & 7.80 & 0.87 & 0.09 & $18.29 \pm 0.09$ & $-16.75 \pm 0.28$ \\
G169.84-07.61 & 6.52 & 0.58 & 0.11 & $18.20 \pm 0.09$ & $-16.30 \pm 0.14$ \\
G170.77-08.51 & 6.84 & 0.86 & 0.16 & $18.55 \pm 0.09$ & $-16.23 \pm 0.15$ \\
G170.88-10.92 & 5.92 & 0.69 & 0.09 & $18.13 \pm 0.18$ & $-16.63 \pm 0.24$ \\
G171.14-17.58 & 7.94 & 0.78 & 0.13 & $18.35 \pm 0.09$ & $-16.92 \pm 0.17$ \\
G171.67-18.05 & 3.92 & 0.94 & 0.14 & $18.51 \pm 0.09$ & $-16.18 \pm 0.15$ \\
G171.80-09.78 & 5.58 & 0.64 & 0.04 & $17.75 \pm 0.09$ & $-17.01 \pm 0.22$ \\
G171.84-05.22 & 7.28 & 0.50 & 0.20 & $18.29 \pm 0.24$ & $-17.27 \pm 0.26$ \\
G172.06-15.21 & 6.54 & 1.01 & 0.08 & $18.31 \pm 0.09$ & $-16.50 \pm 0.16$ \\
G172.57-18.11 & -- & -- & -- & -- & -- \\
G172.92-16.74 & 6.79 & 1.30 & 0.23 & $18.82 \pm 0.09$ & $-16.26 \pm 0.26$ \\
G173.07-16.52 & 6.65 & 1.02 & 0.34 & $18.91 \pm 0.09$ & $-15.80 \pm 0.30$ \\
G173.07-17.89 & 3.77 & 0.81 & 0.11 & $18.34 \pm 0.16$ & $-16.37 \pm 0.19$ \\
G173.89-17.64 & 5.20 & 0.80 & 0.15 & $18.47 \pm 0.09$ & $-16.18 \pm 0.16$ \\
G173.91-16.25 & 6.08 & 1.23 & 0.13 & $18.57 \pm 0.10$ & $-16.16 \pm 0.30$ \\
G175.58-16.60 & 5.54 & 1.69 & 0.18 & $18.79 \pm 0.09$ & $-17.02 \pm 0.22$ \\
\hline
\end{tabular}
\end{table*}

\section{Spectral line fitting toward all sources}\label{sec:lines}

\begin{figure*}
\centering
\includegraphics[width=0.95\linewidth]{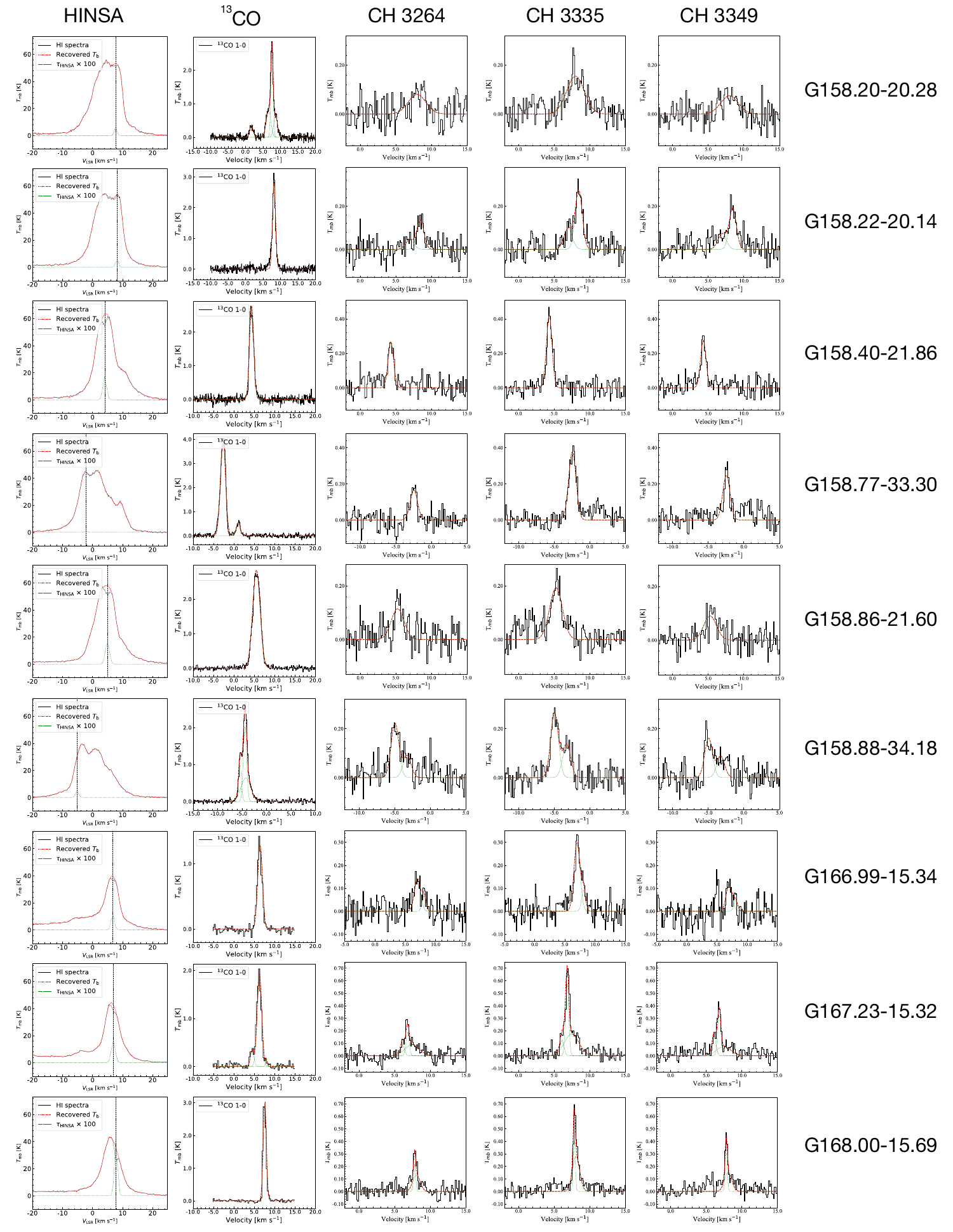}
\caption{From left to right the HINSA, $^{13}$CO, CH 3264 MHz, CH 3335 MHz, and CH 3349 MHz spectra toward our sample of sources as labeled. \label{fig:gsfit1}}
\end{figure*}

\begin{figure*}
\centering
\includegraphics[width=0.95\linewidth]{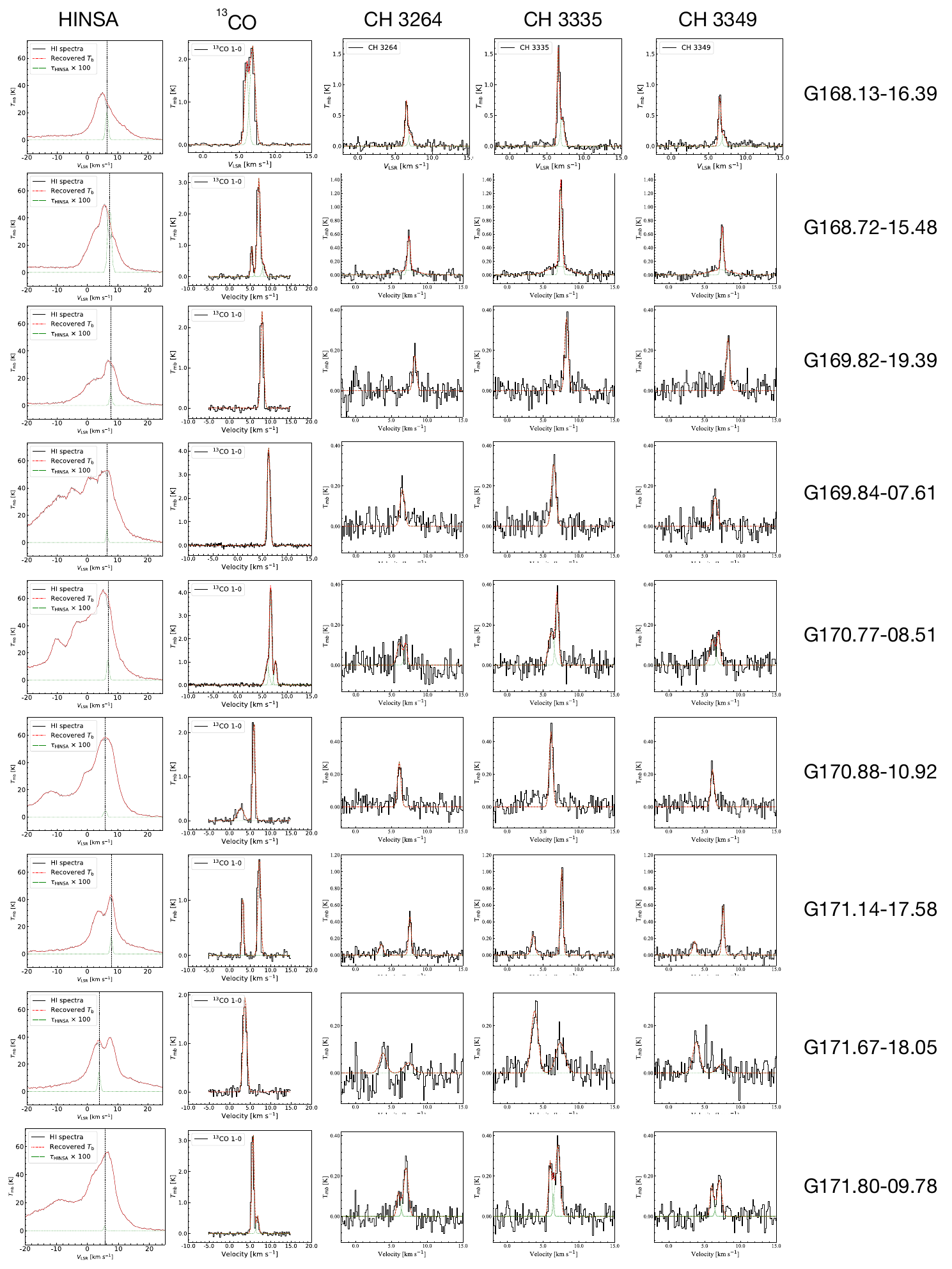}
\caption{Continuation of Fig.~\ref{fig:gsfit1}. \label{fig:gsfit2}}
\end{figure*}

\begin{figure*}
\centering
\includegraphics[width=0.95\linewidth]{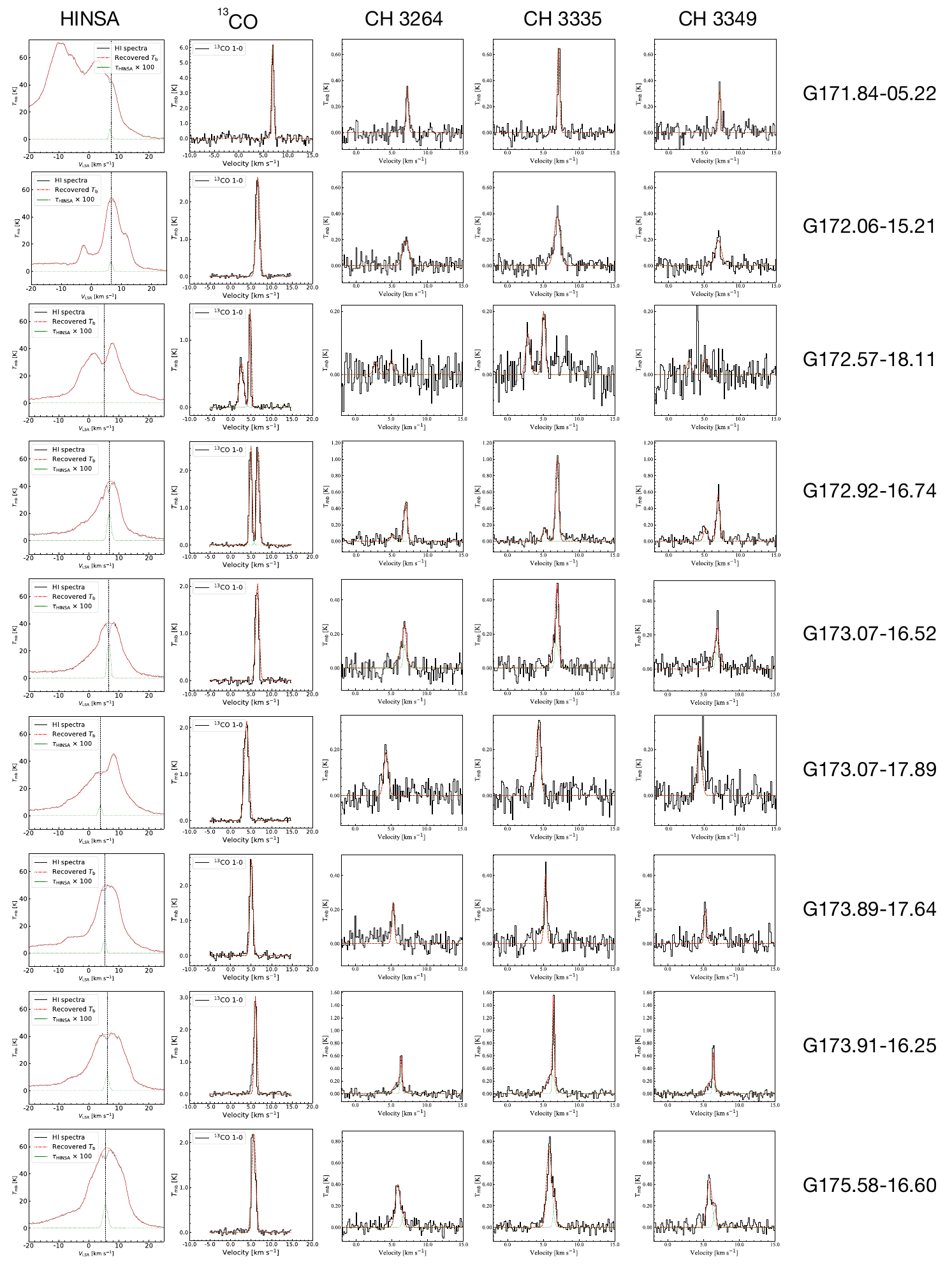}
\caption{Continuation of Fig.~\ref{fig:gsfit2}. \label{fig:gsfit2}}
\end{figure*}

\section{The chemical reactions and reaction rates related to CH formation and destruction}\label{sec:reactions}

Following the reaction networks of CH from \citet{Bialy2015} and \citet{Jacob2023}, and the reaction rates from the latest KInetic Database for Astrochemistry \citep[KIDA,][]{Wakelam2024}, we consider the formation of CH in PGCCs starts from atomic carbon and ionized carbon:
\begin{align}
    \rm C^+ + H_2 \rightarrow &\ \rm CH_2^+ + {\it h\nu}, \label{r1}\\
    \rm C + H_3^+ \rightarrow &\ \rm CH^+ + H_2, \label{r4}\\
    \rm CH^+ + H_2 \rightarrow &\ \rm CH_2^+ + H,
\end{align}
where reaction \ref{r1} is also the formation channel of CH in diffuse clouds \citep{Federman1982}. The recombination reaction between CH$_2^+$ and free electrons will form CH:
\begin{align}
    \rm CH_2^+ + {\it e^-} \rightarrow &\ \rm CH + H, \\
    \rm  \rightarrow &\ \rm C + H_2,\\
    \rm  \rightarrow &\ \rm C + H + H.
\end{align}
In molecular clouds, CH$_2^+$ has a much higher probability of reacting with H$_2$:
\begin{align}
    \rm CH_2^+ + H_2 \rightarrow &\ \rm CH_3^+ + H,
\end{align}
and CH$_3^+$ will react with free electrons to form CH in a branch ratio $\alpha$=0.3:
\begin{align}
    \rm CH_3^+ + {\it e^-} \rightarrow &\ \rm CH + H_2, \\
    \rm  \rightarrow &\ \rm CH + H + H,\\
    \rm  \rightarrow &\ \rm C + H_2 + H,\\
    \rm  \rightarrow &\ \rm CH_2 + H.
\end{align}

The main destruction channels for CH in diffuse clouds are photodissociation, and reactions with H, O, and C$^+$ \citep{Federman1982}. While photodissociation in PGCCs is negligible, the abundant atomic H and O could overwhelm C$^+$ to be the major reactants. We consider here the destruction of CH by atomic H and O:
\begin{align}
    \rm CH + H \rightarrow &\ \rm C + H_2, \label{r2}\\
    \rm CH + O \rightarrow &\ \rm CO + H, \label{r31}\\
    \rm \rightarrow &\ \rm HCO^+ + e, \label{r32}
\end{align}
Considering chemical equilibrium, the abundance of CH can be written as:
\begin{equation}
    f_{\rm CH} = \frac{\alpha(f_{\rm C^+}k_1+f_{\rm H_3^+}f_{\rm C}k_6)}{f_{\rm HI}k_2+f_{\rm O}k_3} = \frac{\alpha f_{\rm C^+}k_1}{f_{\rm HI}k_2+f_{\rm O}k_3}(1+\frac{f_{\rm C}}{f_{\rm C^+}}\frac{k_4}{k_1}f_{\rm H_3^+}),
    \label{eq:fch1}
\end{equation}
where $k_1$, $k_2$, $k_3$, and $k_4$ are the reaction rates of \ref{r1}, \ref{r2}, \ref{r31}+\ref{r32}, and \ref{r4}, respectively. The abundance of H$_3^+$ at high column density should be $f_{\rm H_3^+} \lesssim10^{-8}$ \citep{Indriolo2012}, and the abundance of C$^+$ is comparable or higher than that of atomic C in dense molecular clouds \citep{Beuther2014,Clark2019,Bisbas2023,Bisbas2024}, thus, term $\frac{f_{\rm C}}{f_{\rm C^+}}\frac{k_4}{k_1}f_{\rm H_3^+} \lesssim0.01$. Equation \ref{eq:fch1} can be simplified as:
\begin{equation}
    f_{\rm CH} = \frac{\alpha f_{\rm C^+}k_1}{f_{\rm HI}k_2+f_{\rm O}k_3}.
    \label{eq:fch2}
\end{equation}

Table \ref{table:rates} lists the key reaction rates used in this work.

\begin{table}[H]
\caption{The reaction rates used in this work.}           
\label{table:rates}      
\centering  
\begin{tabular}{rll}
\hline\hline
Reactions & & Rates (cm$^3$~s$^{-1}$)   \\
\hline
${\rm C^+ + H_2 \rightarrow}$ & ${\rm CH_2^+ + {\it h\nu}}$ & $k_1=2\times10^{-16}(T/300)^{-1.3}e^{-23/T}$    \\
${\rm CH + H \rightarrow}$ & ${\rm C + H_2}$ & $k_2=1.24\times10^{-10}(T/300)^{0.26}$    \\
${\rm CH + O \rightarrow}$ & ${\rm CO + H}$ & $k_3=7.0\times10^{-11}$+    \\
${\rm  \rightarrow}$ & ${\rm HCO^+ + e^-}$ & \ \ \ \ \ \ \ \ $2.0\times10^{-11}(T/300)^{0.44}$    \\
${\rm C + H_3^+ \rightarrow}$ & ${\rm CH^+ + H_2}$ & $k_4=2.0\times10^{-9}$    \\
\hline
\end{tabular}
\end{table}

\end{appendix}
%
%
\end{document}